\documentclass[12pt]{article}
\usepackage{epsfig,cite}
\usepackage{axodraw}
\include{epsf}

\def\beq{\begin{equation}}
\def\eeq{\end{equation}}
\def\bear{\begin{eqnarray}}
\def\eear{\end{eqnarray}}

\def\req#1{(\ref{#1})}

\def\gev{{\rm \, Ge\kern-0.125em V}}
\def\tev{{\rm \, Te\kern-0.125em V}}

\def\gappeq{\mathrel{\rlap {\raise.5ex\hbox{$>$}}
{\lower.5ex\hbox{$\sim$}}}}
\def\lappeq{\mathrel{\rlap{\raise.5ex\hbox{$<$}}
{\lower.5ex\hbox{$\sim$}}}}
\def\slash#1{\rlap{\hbox{$\mskip 3 mu /$}}#1}%

\begin{document}
\begin{titlepage}
\pagestyle{empty}
\baselineskip=21pt
\renewcommand{\thefootnote}{\fnsymbol{footnote}}

\rightline{IPPP/10/37} \rightline{DCPT/10/74} \rightline{CERN-PH-TH/2010-107}
\vskip 0.2in
\begin{center}
{\large{\bf Tau-Sneutrino NLSP and Multilepton Signatures at the LHC}}
\end{center}
\begin{center}
\vskip 0.2in
{\bf Terrance Figy}$^1$,
{\bf Krzysztof Rolbiecki}$^2$
and {\bf Yudi Santoso}$^3$
\vskip 0.1in

{\it
$^1${TH Division, CERN, Geneva, Switzerland}\\
$^2${Institute for Particle Physics Phenomenology,
Durham University, Durham DH1 3LE, UK}  \\
$^3${Department of Physics and Astronomy, University of Kansas, Lawrence, \\
KS 66045-7582, USA}
}

\vskip 0.4in
{\bf Abstract}
\end{center}
\baselineskip=18pt \noindent

In models with gravitino as the lightest supersymmetric particle
(LSP), the next to lightest supersymmetric particle (NLSP) can have
a long lifetime and appear stable in collider experiments. We study
the leptonic signatures of such a scenario with tau-sneutrino as the
NLSP, which is realized in the non-universal Higgs masses scenario. We
focus on an interesting trilepton signature with two like-sign taus
and an electron or a muon of opposite sign. The neutralinos and
charginos are quite heavy in the model considered, and the trilepton
signal comes mostly from the slepton-sneutrino production. We
identify the relevant backgrounds, taking into account tau decays,
and devise a set of cuts to optimize this trilepton signal. We
simulate signal and backgrounds at the LHC with 14~TeV
center-of-mass energy. Although the sleptons in this model are
relatively light, $\mathcal{O}(100\ \mathrm{GeV})$, discovery is
more demanding compared to typical neutralino LSP scenarios. The
trilepton signal requires large amount of accumulated data, at least
$\sim 80$~fb$^{-1}$,  at the CM energy of 14~TeV.

\bigskip
\bigskip
\end{titlepage}
\baselineskip=18pt
\renewcommand{\thefootnote}{\arabic{footnote}}

\section{Introduction}

Supersymmetry (SUSY) is one of the candidates for beyond the
standard model (SM) theory that is extensively searched for at
collider experiments (see e.g.~\cite{LHC-ILC}). If supersymmetry does exist at
the weak scale, some supersymmetric particles are expected to be
produced by the Large Hadron Collider (LHC) experiment. To discover
any supersymmetric signal we need a correct theoretical
interpretation of the data, and there have been many studies on this
subject. However, most of these studies assume that the lightest
neutralino is the stable lightest supersymmetric particle
(LSP)~\footnote{The LSP is stable if R-parity is conserved, which we
also assume in this paper.}, motivated by its feasibility as a dark
matter particle~\cite{EHNOS}. Nonetheless, the neutralino is not the
only candidate for dark matter within supersymmetry. It has been
shown that a gravitino LSP can also be a good candidate for dark
matter~\cite{Ellis:2003dn,FengGDM}. In this case, due to its very
weak interactions the gravitino itself would not be seen directly,
while the next-to-lightest supersymmetric particle (NLSP) would
appear as a stable particle at colliders~\footnote{Due to its long
lifetime the NLSP would decay outside of the detectors and appear to
be stable.}.

The gravitino dark matter scenario opens up many new phenomenologies
with various possible NLSPs. If a sneutrino is the NLSP, all other
sparticles that are produced at colliders would quickly cascade
decay to the sneutrino, giving rise to jets and/or leptons along the
way while the sneutrino itself would yield a large missing energy
signature in the detectors. Although this is similar to a neutralino
LSP scenario, the mass spectrum, production rates and branching
ratios are in general different, leading to different
characteristics. Therefore, a dedicated study for the sneutrino NLSP
scenario at colliders is justified. Specifically we look at a
sneutrino NLSP scenario within a supergravity model with
non-universal Higgs masses. A preliminary study on similar model but
within gauge mediated symmetry breaking framework has been performed by
Covi and Kraml in~\cite{Covi:2007xj}, and recently also analyzed by
Katz and Tweedie~\cite{Katz:2009qx}. In this paper, we look into a
detailed analysis, involving Monte Carlo simulation, of a particular
model with tau-sneutrino as the NLSP. We focus on leptonic
channels in order to find distinguishing signatures of the model, and we
study whether the signals in such scenarios can be observed at the LHC.

For hadron colliders, supersymmetric signals can be classified as
jets plus missing energy ($\slash{E_T}$)~\cite{Baer:1995nq}, jets
plus leptons plus missing energy or leptons plus missing energy
without a jet~\cite{Baer:1995va}. Due to the nature of hadron
colliders, signals involving jets are expected to have higher event
rates. However, the standard model backgrounds for these type of
events are also generally larger. On the other hand, although with
relatively small event rates, isolated multilepton plus missing
energy signatures offer relatively clean signals, which in some
cases can be observed above the SM background. For example, a
trilepton signature has been proposed as a promising channel to
discover supersymmetry with a neutralino as the LSP. Motivated by this, and
also because the lightest effectively stable particle in our model
is leptonic, we look at signals with  leptons. We found that the
trilepton signature in our model provides an interesting channel,
which can be used to distinguish it from many other models. However,
search for this signal at the LHC, and hadron colliders in general,
is hindered by the tau identification problem. On the other hand,
inclusive analysis, including jets, shows that the LHC at 14~GeV
should be capable of discovering the new physics beyond the standard
model.

The outline of this paper is as follows. In section~2 we specify our
model and its features, including the sparticle mass spectrum, decay
branching ratios and production rates. In section~3 we explore the
multilepton signatures of our model. Section~4 consists of
discussions on the trilepton signals and backgrounds. In section~5
we show the results of our simulation analysis for the LHC. We
conclude with Section 6.

\section{The Model and Its Features \label{sec:model}}

For our analysis, we take a specific set of parameters in the
Non-Universal Higgs Masses (NUHM) model~\cite{ourNUHM}. The free
parameters in this model are the universal gaugino $m_{1/2}$ and
sfermion masses $m_0$, the trilinear coupling $A_0$ (all three at
the GUT scale); the ratio of the two Higgs vevs $\tan \beta$; the
Higgs mixing parameter $\mu$; and the CP-odd Higgs mass $m_A$ (at
the weak scale). It has been shown that sneutrino NLSP is natural in
NUHM in the sense that it has large parameter space regions allowed
by all known constraints from cosmology, dark matter and particle
physics~\cite{Ellis:2008as}. Our choice of a model corresponds to
NUHM parameters $\tan \beta = 10$, $m_0 = 100$~GeV, $m_{1/2} =
500$~GeV, $A_0 = 0$, $\mu = 600$~GeV and $m_A = 2000$~GeV. It has
tau-sneutrino $\tilde{\nu}_\tau$ as the NLSP and relatively light
slepton masses, while the squarks and gluino are around 1~TeV, still
within the reach of the LHC. The full mass spectrum is listed in
Table~\ref{spectrum-table}. Note that we use a slightly different
value of $m_t$ from the one used in~\cite{Ellis:2008as}.

\vskip 0.15in
\begin{table}[ht]
\renewcommand{\arraystretch}{1.3}
\begin{center}
\begin{tabular}{|l||c|}
\hline
 & Mass [GeV] \\
 \hline
 \hline
$m_{\tilde{\nu}_e}$ & 140.6  \\
 \hline
$m_{\tilde{\nu}_\tau}$ & 90.5  \\
 \hline
$m_{\tilde{e}_L}$ & 161.4  \\
 \hline
$m_{\tilde{\tau}_1}$ & 115.3  \\
 \hline
$m_{\tilde{\chi}^0_1}$ & 206.5  \\
 \hline
$m_{\tilde{\chi}^\pm_1}$ & 396.0  \\
 \hline
$m_{\tilde{\chi}^0_2}$ & 396.1  \\
 \hline
\end{tabular}
\hspace{1cm}
\begin{tabular}{|l||c|}
\hline
 & Mass [GeV] \\
 \hline
 \hline
$m_{\tilde{\chi}^0_3}$ & -617.4  \\
 \hline
$m_{\tilde{\chi}^0_4}$ & 633.0  \\
 \hline
$m_{\tilde{\chi}^\pm_2}$ & 633.5  \\
 \hline
$m_{\tilde{e}_R}$ & 482.7  \\
 \hline
$m_{\tilde{\tau}_2}$ & 459.6  \\
 \hline
$m_{\tilde{t}_1}$ & 723.6  \\
 \hline
$m_{\tilde{t}_2}$ & 994.7  \\
 \hline
$m_{\tilde{b}_1}$ & 956.4  \\
 \hline
$m_{\tilde{b}_2}$ & 1000.9  \\
 \hline
$m_{\tilde{u}_R}$ & 925.6  \\
 \hline
$m_{\tilde{u}_L}$ & 1033.4  \\
 \hline
$m_{\tilde{d}_R}$ & 1012.7  \\
 \hline
$m_{\tilde{d}_L}$ & 1036.5  \\
 \hline
$m_{\tilde{g}}$ & 1176.2  \\
 \hline
\end{tabular}
\hspace{1cm}
\begin{tabular}{|l||c|}
\hline
 & Mass [GeV] \\
 \hline
 \hline
$m_h$ & 115.9  \\
 \hline
$m_H$ &  2000 \\
 \hline
$m_A$ &  2000 \\
 \hline
$m_H^\pm$ & 2002  \\
 \hline
\end{tabular}
\caption{The sparticle and Higgs masses of the model we analyze. We
assume top pole mass $m_t = 172.4$~GeV~\cite{mtop1724} and
$m_b(m_b)^{\overline{\rm MS}} = 4.25$~GeV. The Higgs masses are
calculated using {\tt FeynHiggs}~\cite{fh}.} \label{spectrum-table}
\end{center}
\end{table}
\vskip 0.01in

We assume that the gravitino mass is lower than
$m_{\tilde{\nu}_\tau}$, but we do not need to specify its value as
it is not relevant here. The lifetime of $\tilde{\nu}_\tau$ depends
on the mass gap between gravitino $\widetilde{G}$ and the
tau-sneutrino. However, since the dominant decay channel of
tau-sneutrino is $\tilde{\nu}_\tau \to \widetilde{G} + \nu$, the
tau-sneutrino would still appear as missing energy even if it decays
inside the detector.

Among the lighter sparticles we have the following mass hierarchy
\beq
m_{\tilde{\chi}^0_2}, m_{\tilde{\chi}^\pm_1} >
m_{\tilde{\chi}^0_1} > m_{\tilde{\ell}_L} > m_{\tilde{\nu}_\ell}
> m_{\tilde{\tau}_1} > m_{\tilde{\nu}_\tau} \ ,
\eeq with $\tilde{\nu}_\tau$ as the lightest one. Note that the
first two generations are mass degenerate in the model considered.
Throughout this paper we define $\ell \equiv e, \mu$, whilst
$\ell^\prime \equiv e, \mu, \tau$. The decay modes for these lighter
sparticles are as follows. The chargino can decay as \beq
\tilde{\chi}^\pm_1 \to  \tilde{\chi}^0_1 + W^\pm \ , \quad
\tilde{\ell}_L + \nu_\ell \ , \quad \tilde{\nu}_\ell + \ell \ ,
\quad \tilde{\tau}_1 + \nu_\tau \ , \quad \tilde{\nu}_\tau + \tau \
. \eeq For the neutralinos the decay modes are \beq
\tilde{\chi}^0_{1,2}  \to  \tilde{\ell}_L + \ell \ , \quad
\tilde{\nu}_\ell + \nu_\ell \ , \quad \tilde{\tau}_1 + \tau \ ,
\quad \tilde{\nu}_\tau + \nu_\tau \ , \eeq while for the second
lightest neutralino we have the additional decay mode \beq \tilde{\chi}^0_2
\to  \tilde{\chi}^0_1 + (Z, h) \ , \eeq although with small
branching ratios~\footnote{Note that, $\tilde{\chi}^0_2$ can also
decay through a loop to $\tilde{\chi}^0_1 +
\gamma$~\cite{Ambrosanio:1995az}. However, this is subdominant as
compared to the tree-level two-body decay modes above.}. It is worth
noting that the decay modes for $\tilde{\chi}_1^\pm$ and
$\tilde{\chi}_2^0$ are similar to that in scenarios with
$\tilde{\chi}_1^0$ as the LSP. On the other hand, $\tilde{\ell}_L$
exhibits a completely different decay pattern,
\beq \label{eq:sel-decays} \tilde{\ell}_L \to \tilde{\nu}_\ell +
\bar{f}^\prime + f \ , \quad \tilde{\tau}_1 + \ell + \tau \ , \quad
\tilde{\tau}_1 + \nu_\ell + \nu_\tau \ , \quad \tilde{\nu}_\tau +
\ell + \nu_\tau \ , \quad \tilde{\nu}_\tau + \nu_\ell + \tau \ .
\eeq
Note that only 3-body decay channels are open for the
selectron/smuon. This is because the mass gap between
$\tilde{\ell}_L$ and $\tilde{\nu}_\ell$ is smaller than $m_W$ and
also because of the flavor difference between the selectron and the
NLSP. The decays in Eq.~\req{eq:sel-decays} are mediated by virtual
$W$ ($\tilde{\nu}_\ell \bar{f}^\prime f$), chargino
($\tilde{\nu}_\tau  \nu_\ell  \tau$, $\tilde{\tau}_1 \nu_\ell
\nu_\tau$) or neutralino ($\tilde{\tau}_1  \ell  \tau$,
$\tilde{\nu}_\tau  \ell  \nu_\tau$) exchange. It is interesting to
note that the decay mode $\tilde{\ell}_L \to \tilde{\nu}_\tau  \ell
\nu_\tau$ is highly suppressed because of a destructive interference
between $\tilde{\chi}_1^0$ and $\tilde{\chi}_2^0$-exchange
contributions. Similarly, for the electron-sneutrino we have only
3-body decays
\beq \label{eq:snu-decay}\tilde{\nu}_\ell  \to   \tilde{\tau}_1 +
\nu_\ell + \tau \ , \quad \tilde{\tau}_1 + \ell + \nu_\tau \ , \quad
\tilde{\nu}_\tau + \nu_\ell + \nu_\tau \ , \quad \tilde{\nu}_\tau +
\ell + \tau  \ . \eeq
These decays are mediated by virtual chargino ($\tilde{\tau}_1 \ell
\nu_\tau $, $\tilde{\nu}_\tau \ell \tau$) or neutralino ($
\tilde{\tau}_1 \nu_\ell \tau$, $\tilde{\nu}_\tau \nu_\ell \nu_\tau$)
exchange. The decay width of the sneutrino is highly suppressed with respect
to the left sleptons, see Table~\ref{decay-table}. Heavier selectron mass provides more phase space and
the number of accessible decay modes is significantly larger.
The stau $\tilde{\tau}_1$ can practically~\footnote{Since
the direct decay of stau to gravitino is negligible.} decay only to
the tau sneutrino $\tilde{\nu}_\tau$,
\beq \tilde{\tau}_1 \to \tilde{\nu}_\tau + \bar{f}^\prime + f \ ,
\quad \tilde{\nu}_\tau^* + \nu_\tau + \tau^- \ , \eeq
where the dominant decay mode is mediated via $W$ ($
\tilde{\nu}_\tau \bar{f}^\prime f$) and the other one by chargino
and neutralino. We use {\tt SDecay}~1.3~\cite{sdecay} to calculate
the 2-body decay branching ratios, and {\tt
FeynArts/FormCalc}~\cite{feynarts} package to calculate the 3-body
decay widths~\footnote{The 3-body decays of sleptons and sneutrinos
have also been calculated analytically by Kraml and Nhung
in~\cite{Kraml:2007sx}.}. The branching ratios for the decay
channels with branching ratios $\gappeq 1\%$ are collected in
Table~\ref{decay-table}.  Note that the dominant decay mode for
$\tilde{\nu}_\ell$ is invisible.

\vskip 0.2in

\begin{table}[ht]
\renewcommand{\arraystretch}{1.3}
\begin{center}
\begin{tabular}{|c||c|c|c|c||c|}
\hline
$\tilde{\chi}^+_1 \to$ & $\tilde{\nu}_\tau \tau^+ $ & $\tilde{\nu}_\ell \ell^+$ & $ \tilde{\tau}_1^\ast \nu_\tau$ & $\tilde{\ell}_L^\ast \nu_\ell$ & $\Gamma$ [GeV] \\ \hline
BR [\%] & $18.7$ & $2 \times 15.9 $  & $18.5$ & $2 \times 15.3$ & 7.0 \\ \hline
\end{tabular}

\vskip 0.15in

\begin{tabular}{|c||c|c|c|c||c|}
\hline
$\tilde{\chi}^0_{1,2} \to$ & $\tilde{\nu}_\tau \bar{\nu}_\tau + \mathrm{c.c.}$
& $\tilde{\nu}_\ell \bar{\nu}_\ell  + \mathrm{c.c.}$
& $ \tilde{\tau}_1^\ast \tau^- + \mathrm{c.c.}$
& $\tilde{\ell}_L^\ast \ell^- + \mathrm{c.c.}$
& $\Gamma$ [GeV] \\ \hline
BR ($\tilde{\chi}^0_1$) [\%]&  $2 \times 17.1$ & $4 \times 7.5$ & $2 \times 10.9$ & $4 \times 3.5$ & 0.5 \\ \hline
BR ($\tilde{\chi}^0_2$) [\%]&  $2 \times 9.1$ & $4 \times 7.8$ & $2 \times 9.5$ & $4 \times 7.8$ & 7.0 \\ \hline
\end{tabular}

\vskip 0.15in

\begin{tabular}{|c||c|c|c|c|c|c|c||c|}
\hline
$\tilde{e}^-_L \to$ & $\tilde{\nu}^*_\tau \tau^- \nu_e $ & $\tilde{\nu}_e q_d \bar{q}_u $ & $\tilde{\nu}_e \bar{\nu}_e e^-$ & $ \tilde{\nu}_e \bar{\nu}_\mu \mu^-$ & $\tilde{\nu}_e \bar{\nu}_\tau \tau^-$ & $\tilde{\tau}_1 \tau^+ e^-$ & $\tilde{\tau}_1^\ast \tau^- e^-$ & $\Gamma$ [keV] \\ \hline
BR [\%] & $30.0$ & $ 2 \times 22.0$ & $7.7$ & $7.3$ & $7.3$ & 1.0 & 1.0 & 12 \\ \hline
\end{tabular}

\vskip 0.15in

\begin{tabular}{|c||c|c|c||c|}
\hline
$\tilde{\nu}_\ell \to$ & $\tilde{\nu}_\tau \bar{\nu}_\tau \nu_\ell, \tilde{\nu}_\tau^\ast \nu_\tau \nu_\ell $ & $\tilde{\nu}_\tau \tau^+ \ell^-$ & $\tilde{\tau}_1^\ast \nu_\tau \ell^-$ & $\Gamma$ [keV] \\ \hline
BR [\%] & $70.1$ & $ 21.0$ & $8.4$ &  0.4 \\ \hline
\end{tabular}

\vskip 0.15in

\begin{tabular}{|c||c|c|c||c|}
\hline
$\tilde{\tau}^-_1 \to$ & $\tilde{\nu}_\tau \bar{\nu}_\ell \ell^- $ &  $\tilde{\nu}_\tau \bar{\nu}_\tau \tau^- $ & $\tilde{\nu}_\tau q_d \bar{q}_u $ & $\Gamma$ [keV] \\ \hline
BR [\%] & $2 \times 11.1$ & $11.0$ & $2 \times 33.3$ & 17.2 \\ \hline
\end{tabular}

\caption{Decays and the total widths of $\tilde{\chi}_1^+$,
$\tilde{\chi}_{1,2}^0$, $\tilde{e}_L^-$, $\tilde{\nu}_\ell$ and
$\tilde{\tau}_1$. Only decays with BR $\gappeq 1\%$ are included.
The decay pattern for smuon $\tilde{\mu}^-_L$ is analogous to that
for selectron $\tilde{e}_L^-$. Here $q_u, q_d$ represent $u$-, $c$-
and $d$-, $s$-quarks respectively. Each antiparticle has the same
decay pattern as its corresponding particle. \label{decay-table}}
\end{center}
\end{table}

We calculate the (pair) production rates for the sparticles in our
model at the Tevatron and at the LHC. We assume three center-of-mass
(CM) energies for the LHC: 7~TeV, 10~TeV and 14~TeV. The results are
shown in Table~\ref{prod-table}. Note that the chargino
($\tilde{\chi}_1^\pm$) and neutralinos ($\tilde{\chi}_{1,2}^0$) are
relatively heavy in our model and near the production threshold for
the Tevatron. Note, also, that the squarks and gluinos are not
produced at the Tevatron because of their heavy masses.

\vskip 0.2in
\begin{table}[ht]
\renewcommand{\arraystretch}{1.3}
\begin{center}
\begin{tabular}{|c||c|c|c|c|c|c|c|c|} \hline
\hspace*{-4cm} (a) & $\tilde{\ell}_L^+ \tilde{\ell}_L^-$ &
$\tilde{\nu}_\ell \tilde{\nu}_\ell^*$ & $\tilde{\ell}_L^+
\tilde{\nu}_\ell$ & $\tilde{\ell}_L^- \tilde{\nu}_\ell^*$ &
$\tilde{\tau}_1^+ \tilde{\tau}_1^-$ &  $\tilde{\tau}_1^+
\tilde{\nu}_\tau$ & $\tilde{\tau}_1^- \tilde{\nu}_\tau^*$ &
$\tilde{\nu}_\tau \tilde{\nu}_\tau^*$ \\ \hline Tevatron & 2.9 & 4.7
& 4.4 & 4.4 & 13 & 28 & 28 & 34 \\ \hline 7~TeV LHC & 15 & 26 & 48 &
22 & 57 & 205 & 109 & 153 \\ \hline 10~TeV LHC & 29 & 48 & 86 &  45
& 100 & 344 & 201 & 261 \\ \hline 14~TeV LHC & 51 & 83 & 144 &  81 &
165 & 545 & 339 & 421 \\ \hline
\end{tabular}
\vskip 0.15in
\begin{tabular}{|c||c|c|c|c|c|c|c|} \hline
\hspace*{-4cm} (b)  & $\tilde{\chi}_1^0 \tilde{\chi}_1^0 $ &
$\tilde{\chi}_1^0 \tilde{\chi}_1^-$ & $\tilde{\chi}_1^0
\tilde{\chi}_1^+$ &  $\tilde{\chi}_2^0 \tilde{\chi}_1^-$ &
$\tilde{\chi}_2^0 \tilde{\chi}_1^+$ &  $\tilde{\chi}_2^0
\tilde{\chi}_2^0$ &  $\tilde{\chi}_1^- \tilde{\chi}_1^+$  \\ \hline
Tevatron & 0.03 & 0.002  & 0.002  & 0.07 & 0.07 & 0.002  & 0.17  \\
\hline 7~TeV LHC & 0.3  &  0.03 & 0.11 &  2.9 &  8.2 &  0.19 &  5.5
\\ \hline 10~TeV LHC & 0.7  &  0.08 & 0.26 &  7.8 &  19 &  0.6 &
14.2   \\ \hline 14~TeV LHC & 1.3  &  0.18 & 0.5 &  17 &  38 &  1.4
&  30   \\ \hline
\end{tabular}
\vskip 0.15in
\begin{tabular}{|c||c|c|c|c|c|c|c|c|c|} \hline
\hspace*{-4cm} (c) & $\tilde{q} \tilde{q}^*$ & $\tilde{q}\tilde{q}$
& $\tilde{t}_1 \tilde{t}_1^* $ & $\tilde{g}\tilde{q}$ & $\tilde{g}
\tilde{g}$ & $\tilde{\chi}^0_1 \tilde{q}$ & $\tilde{\chi}^0_2
\tilde{q}$ & $\tilde{\chi}^+_1 \tilde{q}$ & $\tilde{\chi}^-_1
\tilde{q}$  \\ \hline 7~TeV LHC  & 4.4 & 27 & 1.4 & 6.6 & 0.2 & 1.0
& 0.7 & 1.0 & 0.3   \\ \hline 10~TeV LHC  & 34 & 126 & 9.4 & 79 &
4.1 & 3.9 & 3.4 & 5.2 & 2.0   \\ \hline 14~TeV LHC  & 163 & 356 & 43
& 444 & 38 & 14 & 12 & 19 & 7.7   \\ \hline
\end{tabular}

\caption{Cross sections in fb for (a) slepton pair, (b) chargino and
neutralino pair, and (c) squarks and gluino production at the
Tevatron and LHC with CM energies 7, 10 and 14 TeV. The calculation
was done with {\tt Herwig++}~\cite{Herwig}. Note that squarks and
gluino are too heavy to be produced at the Tevatron. Here
$\tilde{q}$ represents the sum over the light squarks $\tilde{u} +
\tilde{d} + \tilde{s} + \tilde{c}$, while $\tilde{\ell}$ can be
either $\tilde{e}$ or $\tilde{\mu}$. \label{prod-table}}
\end{center}
\end{table}


For the light sparticles pair production processes, i.e.\
Table~\ref{prod-table}(a) and (b), we see that at the Tevatron
$\tilde{\nu}_\tau \tilde{\nu}_\tau^*$ (which is invisible) has the
largest cross section due to the light $\tilde{\nu}_\tau$ mass,
followed by $\tilde{\tau}_1^+ \tilde{\nu}_\tau$ and
$\tilde{\tau}_1^- \tilde{\nu}_\tau^*$ (which are the largest visible
channels). For the LHC, which is a proton-proton collider,
$\tilde{\tau}_1^+ \tilde{\nu}_\tau$ has the largest cross section,
followed by $\tilde{\nu}_\tau \tilde{\nu}_\tau^*$ and
$\tilde{\tau}_1^- \tilde{\nu}_\tau^*$. For both colliders gaugino
production is subdominant due to their (relatively) heavy masses,
and in the case of $\tilde{\chi}_1^0$, also, by its bino-dominated
content. As in most models with neutralino LSP (e.g.
SPS1a~\cite{SPS}) $\tilde{\chi}_2^0 \tilde{\chi}_1^\pm$ associated
production is the largest among the gauginos, followed by
$\tilde{\chi}_1^- \tilde{\chi}_1^+$.  For comparison, the
$\tilde{\chi}_2^0 \tilde{\chi}_1^\pm$ production rates for SPS1a at
the LHC is about 900~fb, for 14~TeV CM energy.

As we can see from the table, squarks and gluinos require large
energy because of their heavy masses. At 7~TeV, the production of
squarks and gluinos is negligibly small. At 10~TeV, their total
production rate  is still lower than that of sleptons. At 14~TeV,
the $\tilde{g} \tilde{q}$ becomes important and together with
$\tilde{q} \tilde{q}$ provide promising  channels for SUSY
discovery.

\section{The Leptonic Signatures}

Let us now look at the supersymmetric signals in our model. First,
let us focus on the pure multilepton plus missing energy signals
without  associated jet~\footnote{We will consider inclusive
searches including jets production in the next section.}. These
signals are generated from the production of color singlet sparticles,
i.e.\ the charginos, neutralinos and
sleptons. Thus, we look at chargino pair production
($\tilde{\chi}^\pm_1 \tilde{\chi}^\mp_1$)~\footnote{Same sign
chargino pair can only be produced with some associated
jets~\cite{Alwall:2007ed}.}, neutralino pair production
($\tilde{\chi}^0_i \tilde{\chi}^0_j$),  associated
chargino-neutralino production ($\tilde{\chi}^\pm_1
\tilde{\chi}^0_j$), and slepton pair production
($\tilde{\ell}^{\prime+}_L \tilde{\ell}^{\prime -}_L$,
$\tilde{\ell}^{\prime -}_L \tilde{\nu}_{\ell^\prime}^\ast$,
$\tilde{\ell}^{\prime +}_L \tilde{\nu}_{\ell^\prime}$ and
$\tilde{\nu}^\ast_{\ell^\prime} \tilde{\nu}_{\ell^\prime}$, where
$\ell^\prime = e, \mu, \tau$) as listed in Table~\ref{prod-table}.
From here on, we will implicitly assume the case for the LHC at
14~TeV, unless explicitly stated otherwise.

Let us first look closer at the dominant leptonic decay modes for
sleptons, sneutrinos, the lightest chargino and the second lightest
neutralino. The charged sleptons of the first two generations can
decay directly to $\tilde{\nu}_\tau$ as
\bear
\tilde{\ell}^-_L & \to & \tilde{\nu}_\tau^\ast + \tau^- + \nu_\ell \, ,\nonumber \\
\tilde{\ell}^+_L & \to & \tilde{\nu}_\tau + \tau^+ + \bar{\nu}_\ell
\, . \eear
Note that the decay channels $\tilde{\ell}^- \to \tilde{\nu}_\tau +
\bar{\nu}_\tau + \ell^-,\ \tilde{\nu}_\tau^\ast + \nu_\tau + \ell^-$
are suppressed due to the cancellations mentioned below
Eq.~\req{eq:sel-decays}, in section~\ref{sec:model}. The
selectron/smuon can also decay to the respective sneutrino
\bear
\tilde{\ell}^-_L &\to& \tilde{\nu}_\ell + \ell^{\prime -} + \bar{\nu}_{\ell^\prime} \, , \nonumber \\
\tilde{\ell}^+_L &\to& \tilde{\nu}_\ell^\ast + \ell^{\prime +} +
\nu_{\ell^\prime} \, , \eear
where again $\ell \equiv e, \mu$ and $\ell^\prime \equiv e, \mu,
\tau$; or with  smaller branching ratios to the stau
\bear \tilde{\ell}^\pm_L &\to& \tilde{\tau}_1^- + \tau^+ + \ell^\pm
\, , \hspace{0.5cm}  \tilde{\tau}_1^+ + \tau^- + \ell^\pm \, . \eear
The electron/muon-sneutrino decays mostly invisibly to the
tau-sneutrino and neutrinos. The largest visible decay mode is
\bear
\tilde{\nu}_\ell &\to& \ell^- + \tilde{\nu}_\tau + \tau^+ \, , \nonumber \\
\tilde{\nu}_\ell^\ast &\to& \ell^+ + \tilde{\nu}_\tau^\ast + \tau^- \, . \label{snue-dec-eq}
\eear
They can also decay to stau
\bear
\tilde{\nu}_\ell &\to& \tilde{\tau}_1^+ + \nu_\tau + \ell^- \, , \nonumber \\
\tilde{\nu}_\ell^\ast &\to& \tilde{\tau}_1^- + \bar{\nu}_\tau +
\ell^+ \, . \eear
The leptonic decays of stau are
\bear
\tilde{\tau}^+_1 & \to & \tilde{\nu}_\tau^\ast + \ell^{\prime +} + \nu_{\ell^\prime} \, , \nonumber \\
\tilde{\tau}^-_1 & \to & \tilde{\nu}_\tau + \ell^{\prime -} +
\bar{\nu}_{\ell^\prime} \, . \eear
The second lightest neutralino has much larger production rate (in association with chargino) than the lightest neutralino due to its mostly wino content. It decays as
\bear \tilde{\chi}_2^0 &\to& \tilde{\ell}^{\prime \pm}_L +
\ell^{\prime \mp} \, ,  \hspace{0.5cm} \tilde{\nu}_{\ell^\prime} +
\bar{\nu}_{\ell^\prime} \ , \hspace{0.5cm}
\tilde{\nu}_{\ell^\prime}^\ast + \nu_{\ell^\prime} \, . \eear
The chargino decays as
\bear
\tilde{\chi}_1^+ &\to& \tilde{\nu}_{\ell^\prime} + \ell^{\prime +} \, , \hspace{0.5cm}  \tilde{\ell}^{\prime +}_L + \nu_{\ell^\prime} \, ,  \nonumber \\
\tilde{\chi}_1^- &\to& \tilde{\nu}_\ell^\ast + \ell^- \, ,
\hspace{0.5cm}  \tilde{\ell}^{\prime -}_L + \bar{\nu}_{\ell^\prime}
\, . \eear

We can classify the pure leptonic signals based on the number of the isolated leptons ($e,\mu$ and $\tau$~\footnote{We will consider tau decays in the next section.}) in the final state as follows:
\begin{itemize}
\item[A.] \underline{1 lepton $+ \slash{E_T}$:} \quad
The signals can appear from:
\begin{itemize}
\item[(1)] $\tilde{\tau}_1^- \tilde{\nu}_\tau^\ast$
($\tilde{\tau}_1^+ \tilde{\nu}_\tau$) production with the stau
decays to $\tilde{\nu}_\tau + \bar{\nu}_{\ell^\prime} +
\ell^\prime$, where $\ell^\prime$ could be either $e$, $\mu$ or
$\tau$.
\item[(2)] $\tilde{\ell}_L^- \tilde{\nu}_\ell^\ast$ ($\tilde{\ell}_L^+ \tilde{\nu}_\ell$) production with the sneutrino decaying invisibly as $\tilde{\nu}_\ell \to  \tilde{\nu}_\tau + \nu_\ell + \nu_\tau$, while the selectron/smuon decays as $\tilde{\ell}_L^- \to  \tilde{\nu}_\tau^\ast + \nu_\ell + \tau^-$.
Since the branching ratio for selectron/smuon decay to
$\tilde{\nu}_\tau + \ell + \nu_\tau$ is small, the tau final state
is dominant.
\item[(3)] $\tilde{\chi}^\pm_1 \tilde{\chi}^0_2$ production with
the chargino decays to $\tilde{\nu}_\tau + \tau$ and the neutralino
decays to $\tilde{\nu}_\tau + \nu_\tau$. Again, the tau final state
is dominant.
\end{itemize}
The standard model backgrounds are coming from  direct charged
lepton + neutrino production through $s$-channel $W$-boson exchange,
from single $W$ boson production with the cross section of
20~nb~\cite{Martin:1999ww}, and from $WZ$ with invisible $Z$ with
cross section 3.3~pb~\cite{Frixione:1992pj,Campbell:1999ah}. These
backgrounds are by orders of magnitude larger than the SUSY signals
which are $O(10~{\rm fb})$.

\item[B.] \underline{2 leptons $+ \slash{E_T}$ (dilepton):}  \quad
The SUSY signals can arise from
\begin{itemize}
\item[(1)] $\tilde{\ell}^+_L \tilde{\ell}^-_L$ production where each slepton produces one tau through 3-body decay $\tilde{\ell}_L \to \tilde{\nu}_\tau + \tau + \nu_\ell$.
\item[(2)] $\tilde{\tau}^+_1 \tilde{\tau}^-_1$ production with each stau decaying  through 3-body decay mode $\tilde{\tau}_1 \to \tilde{\nu}_\tau + \nu_{\ell^\prime} + \ell^\prime$ where $\ell^\prime = e, \mu, \tau$.
\item[(3)] $\tilde{\nu}_\ell \tilde{\nu}_\ell^\ast$ pair production with one of the sneutrino decaying as $\tilde{\nu}_\ell  \to \tilde{\nu}_\tau + \ell + \tau$ producing a tau and an electron/muon of opposite signs while the other one decays invisibly as $\tilde{\nu}_\ell  \to   \tilde{\nu}_\tau + \nu_\ell + \nu_\tau$.
\item[(4)] $\tilde{\chi}_1^- \tilde{\chi}_1^+$ pair production with the charginos decaying to $\tau^- \tilde{\nu}_\tau^\ast$ and $\tau^+ \tilde{\nu}_\tau$ respectively.
\end{itemize}
Contributions from neutralino pair production is suppressed by the
small production rate. The SM backgrounds come from direct
production through $\gamma^\ast, Z^\ast$ (Drell-Yan); from single
$Z$ boson production (with cross section of
1.9~nb~\cite{Martin:1999ww}); from $ZZ$ where one $Z$ yields a
neutrino-antineutrino pair while the other $Z$ yields $\ell^{\prime
+} \ell^{\prime -}$ (with cross section of
0.3~pb~\cite{Mele:1990bq,Campbell:1999ah}); and from $W^+ W^-$
production (with cross section of 12.6~pb~\cite{Frixione:1993yp}).
Again the SM backgrounds are much larger than the SUSY signals.

\item[C.] \underline{3 leptons $+ \slash{E_T}$ (trilepton):} \quad
The SUSY signals can come from
\begin{itemize}
\item[(1)] $\tilde{\ell}_L^- \tilde{\nu}_{\ell}^\ast$ ($\tilde{\ell}_L^+ \tilde{\nu}_\ell$) associated production, followed by $\tilde{\ell}_L^- \to \tilde{\nu}_\tau^\ast + \nu_\ell + \tau^-$ and $\tilde{\nu}_\ell^\ast \to \tilde{\nu}_\tau^\ast + \ell^+ + \tau^-$ decays. In this case we have two taus of the same sign and an electron or a muon of the opposite sign.
\item[(2)] $\tilde{\chi}^\pm_1 \tilde{\chi}^0_2$ associated production, with the chargino decays as $\tilde{\chi}^-_1 \to \tau^- + \tilde{\nu}_\tau^\ast$ and the neutralino decays as (a) $\tilde{\chi}^0_2 \to \tilde{\ell}_L^\pm + \ell^\mp$ followed by $\tilde{\ell}_L \to \tilde{\nu}_\tau + \nu_\ell + \tau$, (b) $\tilde{\chi}^0_2 \to \tilde{\tau}^\pm_1 + \tau^\mp$  followed by $\tilde{\tau}_1 \to \tilde{\nu}_\tau + \nu_{\ell^\prime} + \ell^\prime$, or (c) $\tilde{\chi}^0_2 \to \tilde{\nu}_\ell + \nu_\ell$ followed by $\tilde{\nu}_\ell \to \tilde{\nu}_\tau + \tau + \ell$.
\end{itemize}
The SM backgrounds for three leptons are from $WZ$, and
$W\gamma^\ast$~\footnote{At the detector level, there are also some
processes that can mimic trilepton signature such as $ZZ$,
$t\bar{t}$, Drell-Yan and fake leptons~\cite{CDF-trilep}. }. For the
neutralino LSP case, in which the dominant channel is through
$\tilde{\chi}^\pm_1 \tilde{\chi}^0_2$ associated production, this
trilepton signature has been studied thouroughly  and appears to be
a promising channel to discover SUSY~\cite{3lep}. In our scenario,
however, the $\tilde{\chi}^\pm_1 \tilde{\chi}^0_2$ production is
subdominant (and certainly insufficient for the Tevatron), and the
trilepton signals come mostly from $\tilde{\ell}^+_L
\tilde{\nu}_\ell$ and $\tilde{\ell}^-_L \tilde{\nu}_{\ell}^\ast$
production.

It is interesting to notice, however, that for our scenario we can
have two taus of same sign (i.e.\ SF+SS), and that the SM background
for this is expected to be smaller. In this case the SM background
receives contribution from three $W$ bosons production which has a
small cross section.

\item[D.] \underline{4 leptons $+ \slash{E_T}$:} \quad
The SUSY signals can arise from
\begin{itemize}
\item[(1)] $\tilde{\nu}_\ell \tilde{\nu}_\ell^\ast$ production, followed by $\tilde{\nu}_\ell \to \tilde{\nu}_\tau + \tau + \ell$ decays.
\item[(2)] $\tilde{\ell}_L^+ \tilde{\ell}_L^-$ production, with one $\tilde{\ell}_L$ decaying as $\tilde{\ell}_L^- \to \tilde{\nu}_\ell + \ell^{\prime -} + \bar{\nu}_{\ell^\prime}$ (where $\ell^\prime \equiv e, \mu, \tau$) followed by $\tilde{\nu}_\ell \to \tilde{\nu}_\tau + \ell^- + \tau^+$, while the other slepton decays as $\tilde{\ell}_L^+ \to \tilde{\nu}_\tau + \tau^+ + \bar{\nu}_\ell$.
\item[(3)] $\tilde{\chi}^+_1 \tilde{\chi}^-_1$ production, with one of the charginos decaying as $\tilde{\chi}^-_1 \to \tilde{\nu}_\tau^\ast + \tau^-$ and the other one decaying through $\tilde{\chi}^+_1 \to \tilde{\nu}_\ell + \ell^+$ followed by $\tilde{\nu}_\ell \to \tilde{\nu}_\tau + \tau^+ + \ell^-$.
\end{itemize}
The dominant SM background comes from $ZZ$, that has cross section
of 0.12~pb~\cite{Mele:1990bq}. In our scenario Higgs bosons $H$ and
$A$ are quite heavy ($\sim 2$~TeV) and therefore their production at
the LHC would be suppressed. Moreover, the neutralinos and charginos
in our scenario are also relatively heavy. Thus, we do not consider
the same kind of analysis as done in~\cite{4lep-via-H}.

\item[E.] \underline{5 leptons $+ \slash{E_T}$:} \quad The SUSY signals can arise from
\begin{itemize}
\item[(1)] $\tilde{\ell}^+_L \tilde{\nu}_\ell$ ($\tilde{\ell}_L^- \tilde{\nu}_\ell^\ast$) associated production, where the $\tilde{\ell}_L$ decays similarly as in the 4-leptons case producing 3 leptons while the sneutrino decays as $\tilde{\nu}_\ell \to \tilde{\nu}_\tau + \ell + \tau$.

\item[(2)] Again, neutralino-chargino $\tilde{\chi}^0_2 \tilde{\chi}^\pm_1$ associated production gives subdominant contribution. Here the neutralino decays as $\tilde{\chi}_2^0 \to \tilde{\ell}^\prime + \ell^\prime$ followed by $\tilde{\ell}^\prime \to \tilde{\nu}_\tau + \tau + \nu_{\ell^\prime}$~\footnote{Recall that for this specific model the decay $\tilde{\ell}^\prime \to \tilde{\nu}_\tau + \ell + \nu_\tau$ has a very small branching ratio. Having this decay channel available, we would have 5-lepton signature with only one tau.}, while the chargino decays as $\tilde{\chi}_1^\pm \to \tilde{\nu}_\ell + \ell$ followed by $\tilde{\nu}_\ell \to \tilde{\nu}_\tau + \ell + \tau$.
\end{itemize}
The SM backgrounds are from $WZZ$~\cite{tripleWZ}, $WZ\gamma^\ast$
and $W\gamma^\ast \gamma^\ast$. Note that even though the SUSY
5-lepton signal has a small rate $\mathcal{O}(0.1~{\rm fb})$,
suppressed by branching ratios of $\tilde{\ell}_L$ and
$\tilde{\nu}_\ell$ decays, the SM background is also small. Thus
this might also be an interesting channel to look at. The question,
however, is how much luminosity would be needed to receive enough
significance.
\end{itemize}

\section{The Trilepton Signals and Backgrounds}

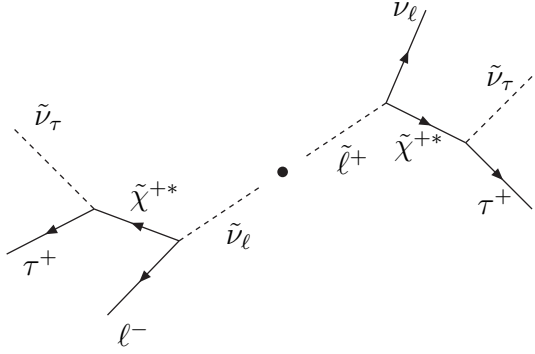
\begin{figure}
\begin{center}
\begin{picture}(220,132)(0,0)
  \DashLine(135,75)(165,95){2}  \Text(153,74)[]{$\tilde{\ell}^+$}
  \ArrowLine(165,95)(195,80) \Text(178,79)[]{$\tilde{\chi}^{+ \ast}$}
  \ArrowLine(165,95)(180,130)  \Text(173,130)[]{$\nu_\ell$}
  \DashLine(195,80)(220,105){2}  \Text(209,105)[]{$\tilde{\nu}_\tau$}
  \ArrowLine(195,80)(220,55)  \Text(207,58)[]{$\tau^+$}
  \Vertex(126,69){2}
  \DashLine(117,63)(87,43){2}  \Text(110,45)[]{$\tilde{\nu}_\ell$}
  \ArrowLine(87,43)(55,55) \Text(78,60)[]{$\tilde{\chi}^{+ \ast}$}
  \ArrowLine(87,43)(60,15)  \Text(70,8)[]{$\ell^-$}
  \DashLine(55,55)(25,85){2}  \Text(38,90)[]{$\tilde{\nu}_\tau$}
  \ArrowLine(55,55)(22,38)  \Text(35,35)[]{$\tau^+$}
\end{picture}
\end{center}
\vskip -0.2in
\caption{Example of trilepton signature from slepton-sneutrino associated production. \label{fig:diagram}}
\end{figure}

As mentioned in the previous section, the trilepton signature with a
pair of like-sign taus is particularly interesting. The signals that
we are looking for are $\tau^+ \tau^+ (e,\mu)^-$ and $\tau^- \tau^-
(e,\mu)^+$, which in our SUSY scenario arise mainly from
slepton-sneutrino associated production followed by cascade decays
(illustrated in Fig.~\ref{fig:diagram}). If the taus and their
charges can be identified in the detectors then this would provide
us with an excellent supersymmetric signal with a distinctly larger
cross section than the standard model background. In the SM this
signature can be mimicked, primarily, through the production and
decay of three $W$--bosons:
\begin{eqnarray}
pp\to W^{+} W^{+} W^{-} \to \tau^{+} \nu_{\tau} \tau^{+} \nu_{\tau} \ell^{-} \bar{\nu}_{\ell} \, ,
\end{eqnarray}
and
\begin{eqnarray}
pp\to W^{-} W^{-} W^{+} \to \tau^{-} \bar{\nu}_{\tau} \tau^{-} \bar{\nu}_{\tau} \ell^{+} \nu_{\ell} \, ,
\end{eqnarray}
respectively.

In reality, however, taus decay quickly inside the detectors,
producing either leptons (i.e.\ $e$, $\mu$ and neutrinos) or jets
(plus tau-neutrino). Tau identification could present a problem,
especially for a hadron collider such as the LHC with high jet multiplicities.
If a tau decays to $e/\mu$ then it would be difficult
to distinguish it from the electrons/muons produced by other
processes. On the other hand it is not easy, although not
impossible, to identify jets that are coming from
taus~\cite{Atlas-LHC,CMS-TDR}. Let us recapitulate on the signals as
seen by the detectors:
\begin{itemize}
\item[(a)] $e^\pm \mu^\pm (e,\mu)^\mp$,
\item[(b)] $e^\pm e^\pm e^\mp$, $\mu^\pm \mu^\pm \mu^\mp$,
\item[(c)] $e^\pm e^\pm \mu^\mp$, $\mu^\pm \mu^\pm e^\mp$,
\item[(d)] $\tau_h^\pm (e^\pm \mu^\mp, \mu^\pm e^\mp)$,
\item[(e)] $\tau_h^\pm (e^\pm e^\mp, \mu^\pm \mu^\mp)$,
\item[(f)] $\tau_h^\pm \tau_h^\pm (e,\mu)^\mp$.
\end{itemize}
Here $\tau_h$ represent a hadronic tau. The ratios are about
\begin{equation}
(a):(b):(c):(d):(e):(f) \simeq 6 : 3 : 3 : 22 : 22 : 42
\end{equation}
At this level there is another SM background from the following process:
\begin{equation}
pp \to W^+ Z \to \tau^+ \tau^+ \tau^- \nu_\tau
\end{equation}
with the taus decaying either leptonically or hadronically. In
addition, there is a background from $WWW$ which can produce the
leptonic signals directly without going through any tau. Signals
(a), (b) and (e) also receive backgrounds from $WZ$ and $W
\gamma^\ast$ which produce $e/\mu$ directly. Thus, the interesting
signals to look at are (c), (d) and (f). Signal (c) provide a clear
signature, but is suppressed by the branching ratios. Although (d)
is quite interesting, the signals might be overwhelmed by fake taus.
Therefore we concentrate on (f) in our analysis, where we look for
two tau-jets of same sign and a muon/electron of the opposite sign.
From here on, we will always mean hadronic tau ($\tau_h$) when we
say tau ($\tau$), unless  explicitly stated otherwise.

At the simulation level, we need to consider some additional
backgrounds. This is due to the fact that there could be some
leptons or jets that do not pass the selection criteria, resulting
in a different signature. For example, we can have $ZZ$ with each
$Z$ decaying to a pair of taus, then two taus decay hadronically,
one tau leptonically while the other one is missing. Thus, for
$\tau^+ \tau^+ \mu^-$ signal, we need to include $Z$--pair, top-pair, and single top -- $W$
associated production as well.

The detection of hadronic tau is also not straightforward. Full
analysis would require detector simulation and tau reconstruction,
which are beyond the scope of our paper. To take tau identification
problem into account we can make an estimate by attaching a
detection efficiency factor to each hadronic tau, $0 < \epsilon_h < 1$.
However, to be more precise this factor should be taken as a
function of transverse momentum and
pseudo-rapidity~\cite{CMS-tau,ATLAS-tau}. Note that this factor
affects both the signal and the background. For this reason we do
not include this factor in the histograms shown in the next section
below. In addition, the tau charge should also be identified
correctly for our case. Charge identification is expected to become
worse for larger tau momentum, although it should not be impossible
for the interesting range in our model at the LHC~\cite{Atlas-LHC}.
This charge identification can be used to eliminate some background events arising
from $t \bar{t}$, but not entirely.

At the detector level, there could be additional backgrounds from
jets that are misidentified as tau's, i.e.\ fake tau signals. For
example, the $W j j$ which has a much larger production
rate~\cite{W2j} can be problematic. The rejection rate of fake taus
depends on detector's capability and is correlated to the tau
identification efficiency~\cite{Atlas-LHC}. If we assume (effective)
rejection factor of 500, with $e^+ \nu_e jj$ cross section of
670~pb, we obtain 2.4~fb of fake tau background, which is comparable
to the SUSY signal. We notice that the missing transverse energy for
this background is below 200~GeV.

On the other hand, hadron colliders, such as the LHC, produce many
jets in both SUSY and SM processes. By looking at trilepton plus
any number of jets we would obtain more signal events. We start
with inclusive search of $\tau^\pm \tau^\pm \mu^\mp + nj$, where $n
= 0,1,2, \ldots$, and then employ cuts to reduce the backgrounds.
Our SUSY signal now consists of slepton-sneutrino,
chargino-neutralino and SUSY QCD. In SUSY QCD, squarks and gluinos
are produced and cascade decay to the tau-sneutrino NLSP. We found
that this SUSY QCD contribution gives large transverse energy to the
final states, due to the big gap between the squark sector and the
slepton sector in our model. Thus this can be used as first evidence
of new physics beyond the standard model.

\section{Analysis and Results at the LHC}

In this section we study the inclusive trilepton signals at the LHC
for a  CM energy of $14$~TeV. The inclusive SUSY signal has been
generated with Monte Carlo program {\tt
Herwig++~2.4.2}~\cite{Herwig}. We have included all sparticle pair
production processes and all possible $2$--body and $3$--body
sparticle decays in the {\tt Herwig++} simulations. All background
processes have been simulated with Monte Carlo program {\tt
SHERPA~1.2.0} ~\cite{sherpa}. For the {\tt SHERPA} simulations we
have used {\tt COMIX}~\cite{comix} to compute the hard  matrix
elements.

After generating events, we apply the following selection criteria:
\begin{enumerate}
\item Jets reconstructed according the anti-$k_{T}$ algorithm with $D=0.7$~\cite{antikTalg} which are required to have
\begin{eqnarray}
p_{T}^{j} > 20~{\rm GeV}\;, \quad \quad |\eta_{j}|<4.5 \; .
\end{eqnarray}
\item $N_{\mu}=1$ : Isolated muons with $R_{\mu,j}>0.7$ .
\item $N_{\tau}=2$ : Isolated like sign taus with $R_{\tau,j}>0.7$ .
\item The hardest lepton is required to have:
\begin{eqnarray}
p_{T}^{\ell} > 10~{\rm GeV}\;, \quad \quad |\eta_{\ell}|<2.5 \; .
\end{eqnarray}
\item The two hardest taus in the event are required to have:
\begin{eqnarray}
p_{T}^{\tau_h} > 15~{\rm GeV}\;, \quad \quad |\eta_{\tau_h}|<2.5 \; .
\end{eqnarray}
\item Leptons and taus are required to be isolated with
\begin{eqnarray}
R_{\ell,\tau_h} > 0.4 \; \quad \quad R_{\tau_h,\tau_h}>0.4 \;.
\end{eqnarray}
\end{enumerate}
These form our basic cuts.
We have used {\tt Rivet~1.2.1}~\cite{rivet} and {\tt
FastJet~2.4.2}~\cite{fastjet} in order to analyze events according
to our prescribed selection criteria.

\begin{figure}[htbp]
\begin{center}
\includegraphics[width=0.48\linewidth]{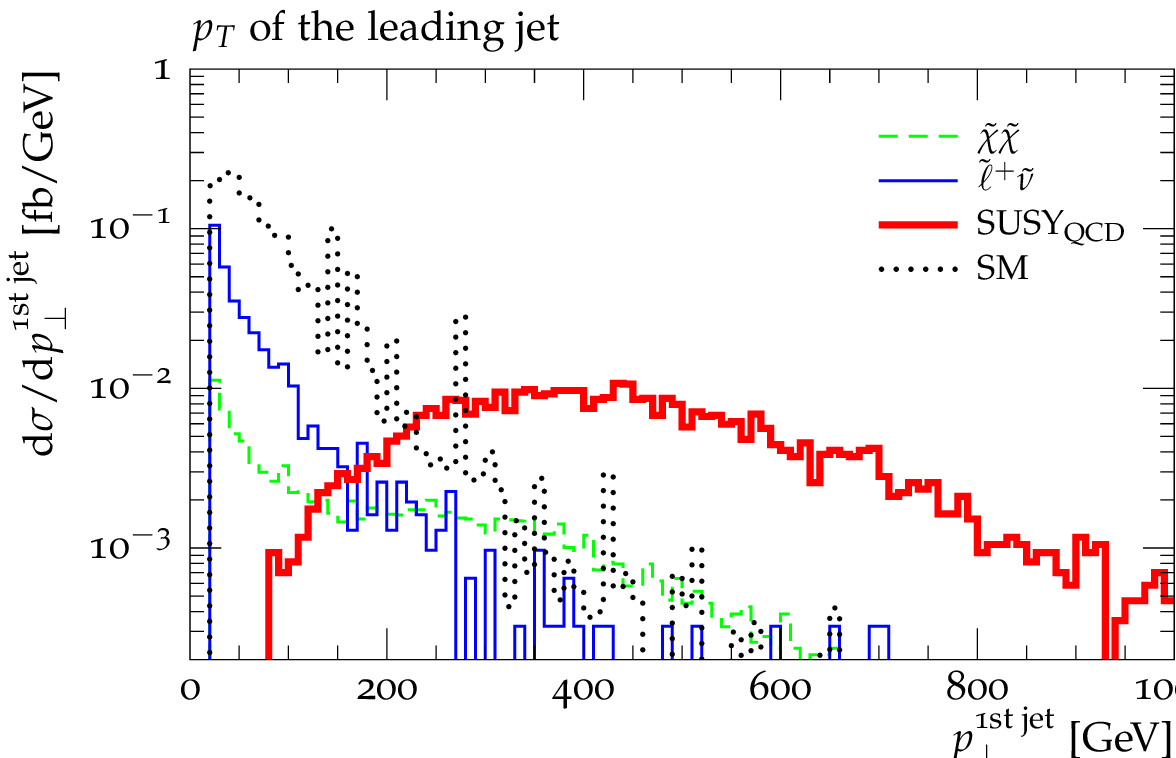}
\includegraphics[width=0.48\linewidth]{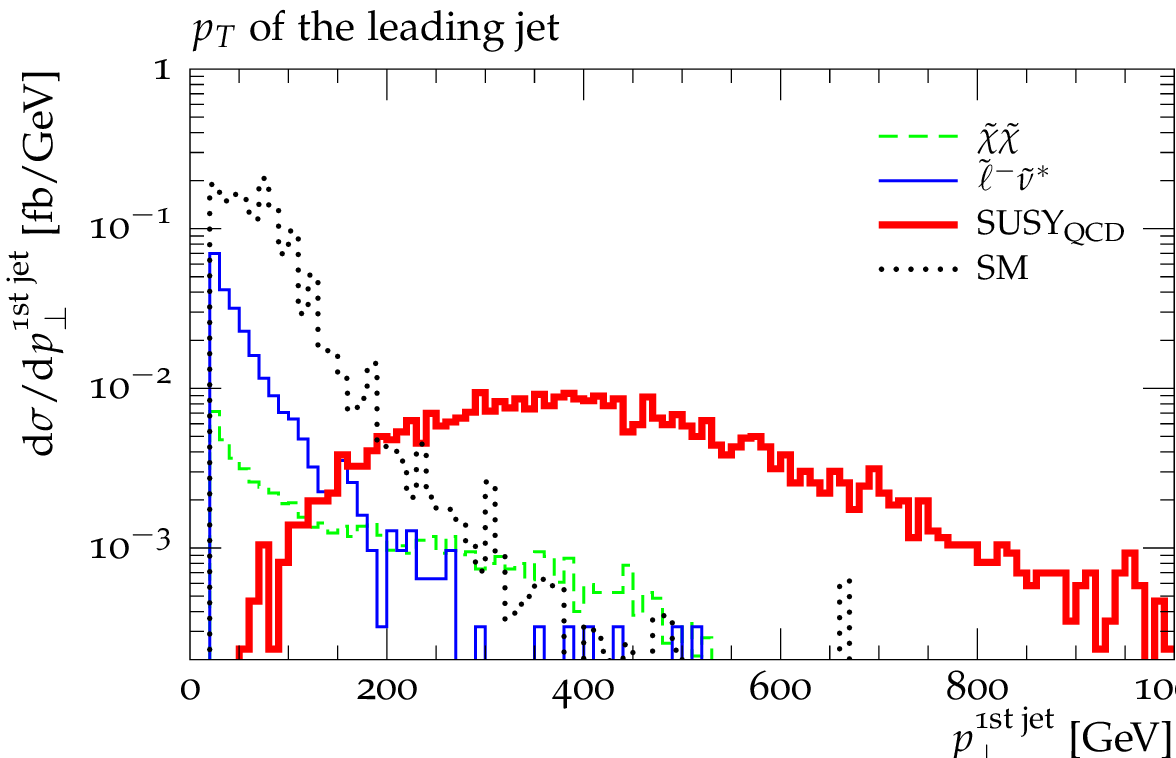}
\hfill (a) \hspace{7cm} (b) \hfill
\caption{The distribution of the hardest jet transverse momentum,
$p_{T}^{\rm j1}$ for (a) $\tau^+ \tau^+ \mu^- +$ jets and (b)
$\tau^- \tau^- \mu^+ +$ jets for SUSY signals and SM background. }\label{Fig:PT-jet-basic}
\end{center}
\end{figure}

In Fig.~\ref{Fig:PT-jet-basic} we show the transverse momentum
distribution for the hardest jet, $p_{T}^{j1}$, after the basic
cuts. It is obvious that the distribution at large $p_T$ is
dominated by contributions from SUSY QCD, i.e.\ from production of squarks and
gluinos. Similarly for the second, third and fourth
hardest jets. This would provide a clear signal of new physics
beyond the standard model. Note that this feature should also be
found for signals with any number of leptons in our scenario, and
also for other scenarios in which squarks and gluino are much
heavier than the LSP. Thus, although high $p_T$ jets indicate new
physics, it is not a unique feature of our model.

We then apply optimized cuts to enhance the signal-to-background
ratio. At this point we have two branches of analysis. The main
branch is focusing on the leptonic features of the SUSY signal, with
the following set of cuts:
\begin{itemize}
\item[A.]
\begin{enumerate}
\item Veto on $b$--jets and more than one jet, i.e.\  $N_{b} = 0$ and $ N_{j} \le 1$.
\item Cut on the transverse momentum of the hardest jet in the event above 200~GeV: i.e. we require $20~{\rm GeV} < p_{T}^{\rm j1} \le ~200~{\rm GeV} $ .
\item Require $m_{\rm min }(\mu, \tau) = {\rm min}(m(\mu_{1},\tau_{1}),m(\mu_{1},\tau_{2})) < 55~ {\rm GeV}$ and $\phi( \mu, \slash{p_T}) \ge 1.5$ rads.
\end{enumerate}
\end{itemize}
We call this set opt.A for short. It optimizes the SUSY EW signal.
For the side branch analysis, we have the following set of cuts
(opt.B) which is designed to promote the high-$p_T$ jets of SUSY QCD
signal~\cite{Baer:2007ya,Baer:2008kc}.
\begin{itemize}
\item[B.]
\begin{enumerate}
\item We require $N_{j} \ge 2$ and $p_{T}^{\rm j1} \ge ~200~{\rm GeV} $, and
\item $A_{T}^{\prime} = \sum_{i={\rm leptons, jets}}{E_{T}^{i}}  \ge 300~ {\rm GeV}$ .
\end{enumerate}
\end{itemize}

The results of our simulations are tabulated in
Table~\ref{Tab:trilep1} for the $\ell = \mu$ case. The $\ell = e$
case is similar to the  $\mu$ case, hence is not shown here.
Backgrounds for the case of the $\tau^{+} \tau^{+} \ell^{-}$
signature include $\tau^{+} \nu_{\tau} \tau^{+} \nu_{\tau} \mu^{-}
\bar{\nu}_{\mu}$ denoted simply as $W^{+}W^{+}W^{-}$, $\tau^{+}
\tau^{-} \tau^{+} \nu_{\tau}$ denoted as $ZW^{+}$, $\tau^{+}
\tau^{-} \tau^{+} \tau^{-}$ denoted as $ZZ$, $W^{+}[ \to \tau^{+}
\nu_{\tau} ] W^{-}[ \to \mu^{-} \bar{\nu}_{\mu}] b \bar{b} \tau^{+}
\tau^{-}$ denoted as $t \bar{t} Z$, $t[ \to b \tau^{+} \nu_{\tau}]
\bar{t} [\to \bar{b} \mu^{-} \bar{\nu}_{\mu}] W^{+} [\to \tau^{+}
\nu_{\tau}] $ denoted as $ t  \bar{t} W^{+}$, $W^{+} [\to \tau^{+}
\nu_{\tau}] W^{-} [\to \mu^{-} \bar{\nu}_{\mu}] b \bar{b}$ denoted
as $W^{+} W^{-} b \bar{b}$, and $\mu^{-} \bar{\nu}_{\mu} \tau^{+}
\nu_{\tau} \tau^{+} \tau^{-}$ denoted as $ZW^{+}W^{-}$. Note that
the backgrounds from top pair production, single top -- $W$ boson
associated production are already included in $W^+W^-b \bar{b}$. We
have similar backgrounds for the $\tau^{-} \tau^{-} \ell^{+}$, but
with the charges conjugated.

In opt.A, the veto on jets helps to reduce the SM QCD
background, in particular $t\bar{t}$, although it also suppresses
the SUSY QCD signal.  The cuts on $m_{\rm min }(\mu, \tau)$ and
$\phi( \mu, \slash{p}_{T})$ are used to suppress backgrounds from
$ZZ$ and $ZW$. As we can see from the table, the SUSY signal is now
comparable to the SM background. In total, it is greater than the
background for both $\tau^+ \tau^+ \mu^-$ and $\tau^- \tau^- \mu^+$
cases, but we obtain an improved result for the $\tau^+ \tau^+ \mu^-$
case. With opt.B, on the other hand, SUSY EW signal is suppressed
due to the low jet multiplicity. We see that after the optimization
we obtain a SUSY QCD signal significantly higher than the backgrounds.


\begin{table*}
\renewcommand{\arraystretch}{1.3}
\begin{center}
\begin{tabular}{|l||c|c|c|}
\hline
 $ \tau^{+}  \tau^{+} \mu^{-}$ & $\sigma_{\rm basic}$[fb] & $\sigma_{\rm opt A}$ [fb]  & $\sigma_{\rm opt B}$[fb]  \\
\hline
\parbox{25mm}{\hspace{0.1in} Susy EW}  & 3.55 & 1.78 & 0.0828    \\
\parbox{25mm}{\hspace{0.1in} Susy QCD} & 4.09 & 0.00 & 3.73 \\
\parbox{25mm}{\hspace{0.1in} Susy $\chi \chi$} & 1.83 & 0.0986 & 0.322 \\
\parbox{25mm}{\hspace{0.1in} $ZW^{+}$}  & 4.80 & 0.829  & 0.200          \\
\parbox{25mm}{\hspace{0.1in} $ZZ$}   & 1.80 & 0.172 & 0.0164       \\
\parbox{25mm}{\hspace{0.1in} $W^{+} W^{-} b \bar{b}$}  & 10.4 & 0.0390 & 0.285     \\
\parbox{25mm}{\hspace{0.1in} $t \bar{t} W^{+}$}  & 0.0506 & $5.81 \times 10^{-5}$ & 0.00289   \\
\parbox{25mm}{\hspace{0.1in} $t \bar{t} Z$}  & 0.127 & $3.50 \times 10^{-5}$ & 0.00642   \\
\parbox{25mm}{\hspace{0.1in} $W^{+}W^{+}W^{-}$}   & 0.0728 & 0.0117 & 0.00423  \\
\parbox{25mm}{\hspace{0.1in} $Z W^{+}W^{-}$}  & 0.0348 & 0.00453 & 0.00232  \\
\hline
$ \tau^{-}  \tau^{-} \mu^{+}$ & $\sigma_{\rm basic}$ [fb] & $\sigma_{\rm opt A}$ [fb]  & $\sigma_{\rm opt B}$[fb]  \\
\hline
\parbox{25mm}{\hspace{0.1in} Susy EW }  & 2.46 & 1.24 & 0.0523   \\
\parbox{25mm}{\hspace{0.1in} Susy QCD}  & 3.51 & 0.00150 & 3.18   \\
\parbox{25mm}{\hspace{0.1in} Susy $\chi \chi$}  & 0.676 & 0.0676 & 0.203 \\
\parbox{25mm}{\hspace{0.1in} $ZW^{-}$}   & 3.64 & 0.633 & 0.0927  \\
\parbox{25mm}{\hspace{0.1in} $ZZ$}    & 1.78 & 0.161 & 0.0161    \\
\parbox{25mm}{\hspace{0.1in} $W^{+} W^{-} b \bar{b}$}   & 9.07 & 0.0204 & 0.0529   \\
\parbox{25mm}{\hspace{0.1in} $t \bar{t} W^{-}$}   & 0.0305 & $5.02  \times 10^{-5}$ & 0.00137  \\
\parbox{25mm}{\hspace{0.1in} $t \bar{t} Z$}  & 0.135 & $5.36 \times 10^{-5}$ & 0.00571   \\
\parbox{25mm}{\hspace{0.1in} $W^{+}W^{-}W^{-}$}   & 0.0498 & 0.0106 & 0.00299   \\
\parbox{25mm}{\hspace{0.1in} $Z W^{+}W^{-}$}    & 0.0333 & 0.00480 & 0.00236  \\
\hline
\end{tabular}
\parbox{0.9\textwidth}{\caption{Generation characteristics for
$pp \to \mu^{-} \tau_h^{+} \tau_h^{+} + \slash{E}_{T}$ and $pp \to
\mu^{+} \tau_h^{-} \tau_h^{-} + \slash{E}_{T}$. Tau detection
efficiency is not included. }
    \label{Tab:trilep1}}
\end{center}
\end{table*}

\begin{figure}[htbp]
\begin{center}
\includegraphics[width=0.48\linewidth]{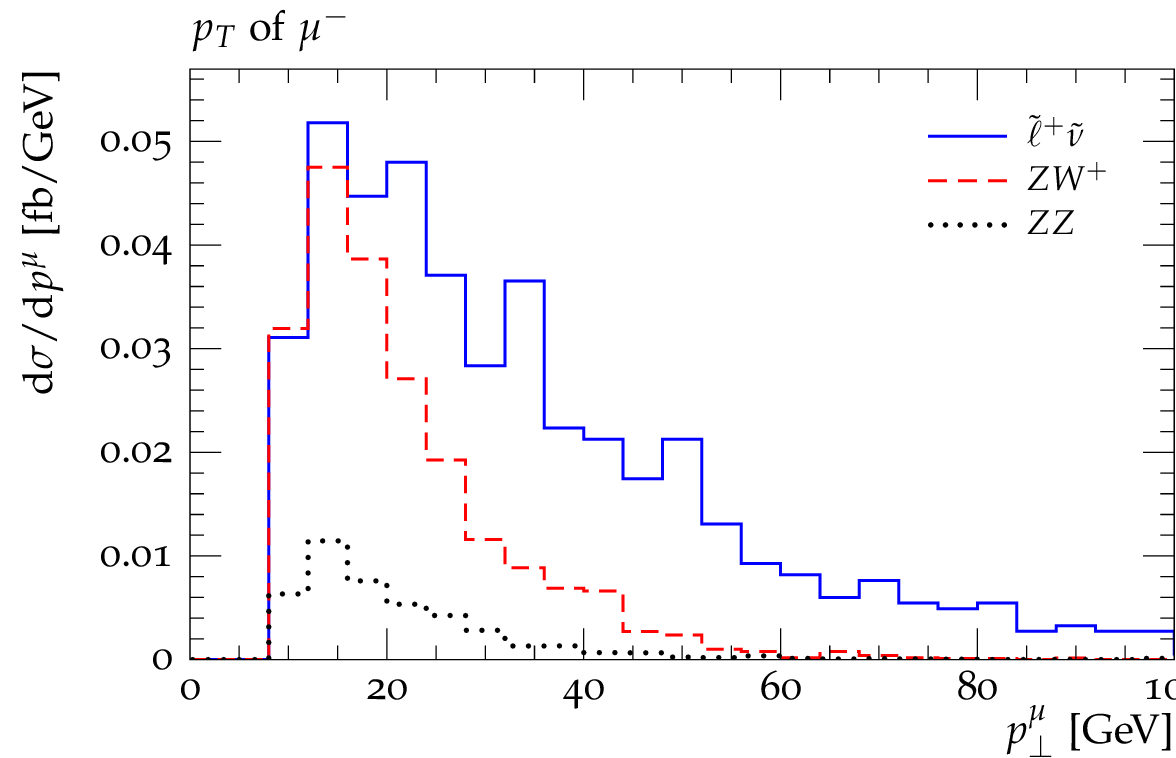}
\includegraphics[width=0.48\linewidth]{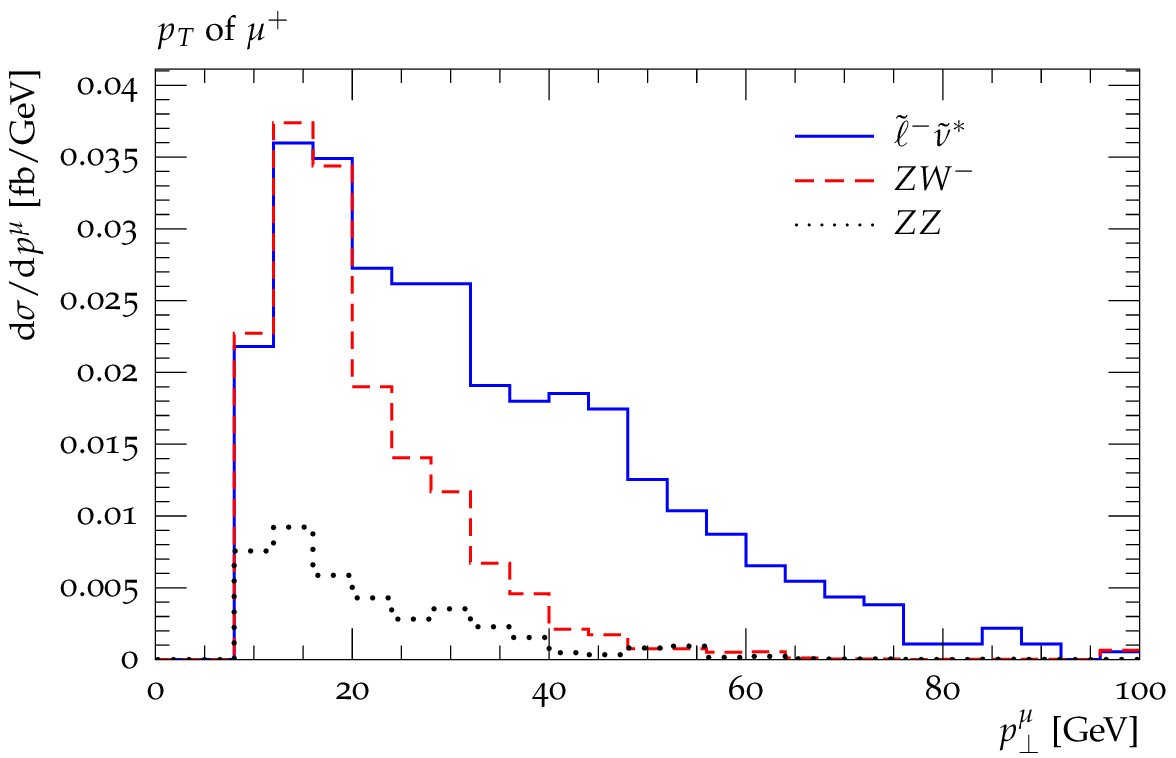}
\hfill (a) \hspace{7cm} (b) \hfill
\caption{The distribution in the (a) $\mu^{-}$ and (b) $\mu^{+}$ transverse momenta, $p_{T}^{\mu^{-}}$ and $p_{T}^{\mu^{+}}$ respectively. Channels giving negligible contribution are not shown here. }\label{Fig:PT-mu}
\end{center}
\end{figure}

We now focus our discussion on the main analysis (i.e.\ opt.A). In
Fig.~\ref{Fig:PT-mu} we show the muon transverse momentum
distribution $p_T^\mu$ after the optimized cuts. The largest
background comes from $ZW$. The shape of $ZW$ is following that of
SUSY EW, but is is softer. However, the signal distribution is
larger for smaller $p_T^\mu$, decreasing rapidly with increasing
$p_T^\mu$. Therefore there is less incentive to optimize the cut on
$p_T^\mu$.

\begin{figure}[htbp]
\begin{center}
\includegraphics[width=0.48\linewidth]{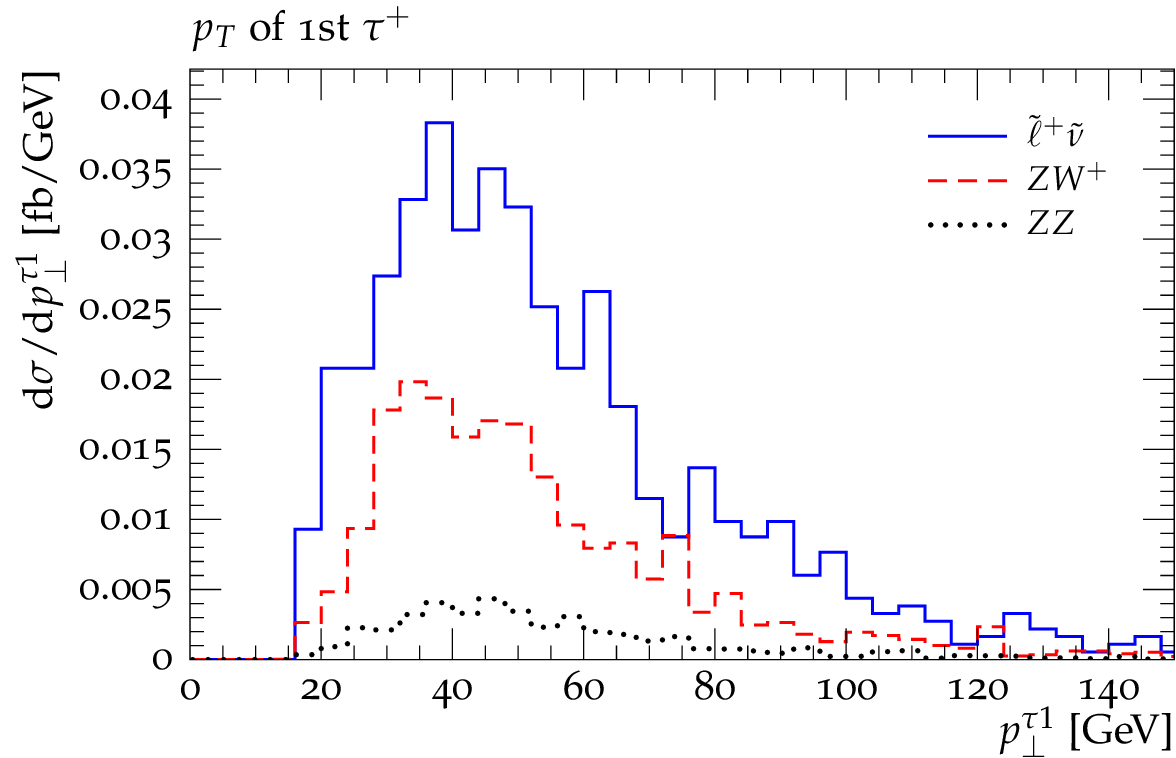}
\includegraphics[width=0.48\linewidth]{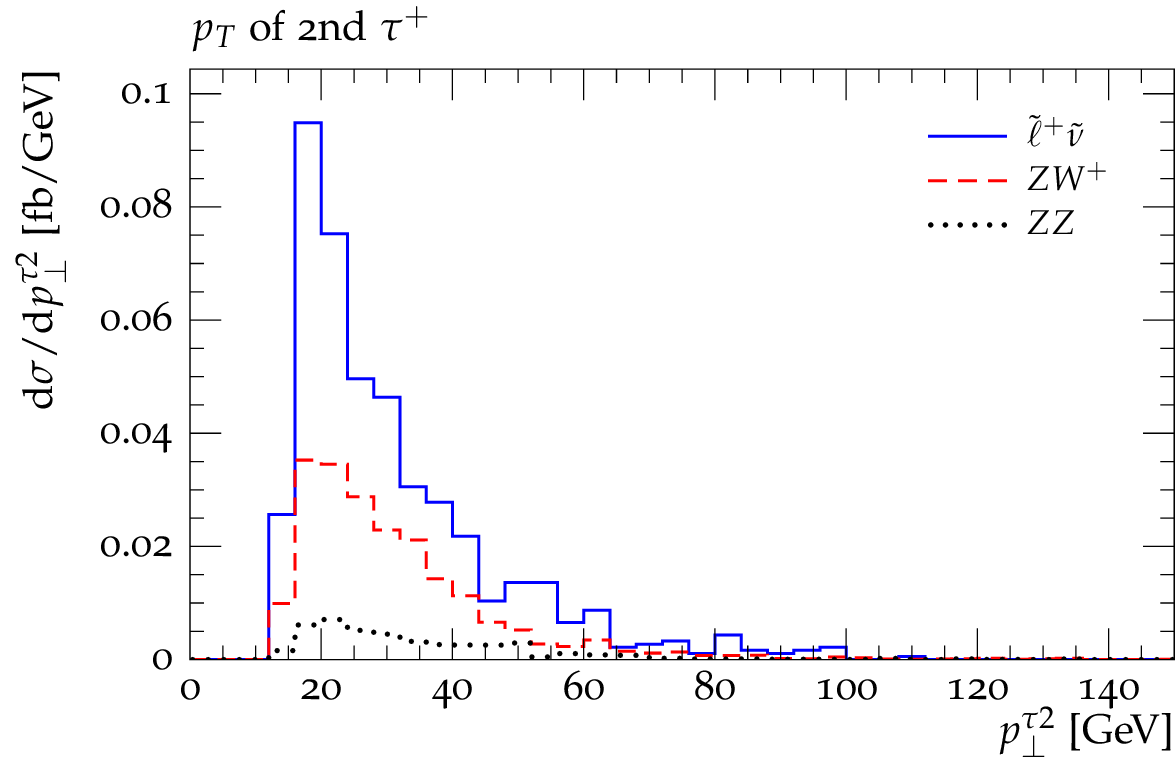}
\hfill (a) \hspace{7cm} (b) \hfill
\vskip 0.1in
\includegraphics[width=0.48\linewidth]{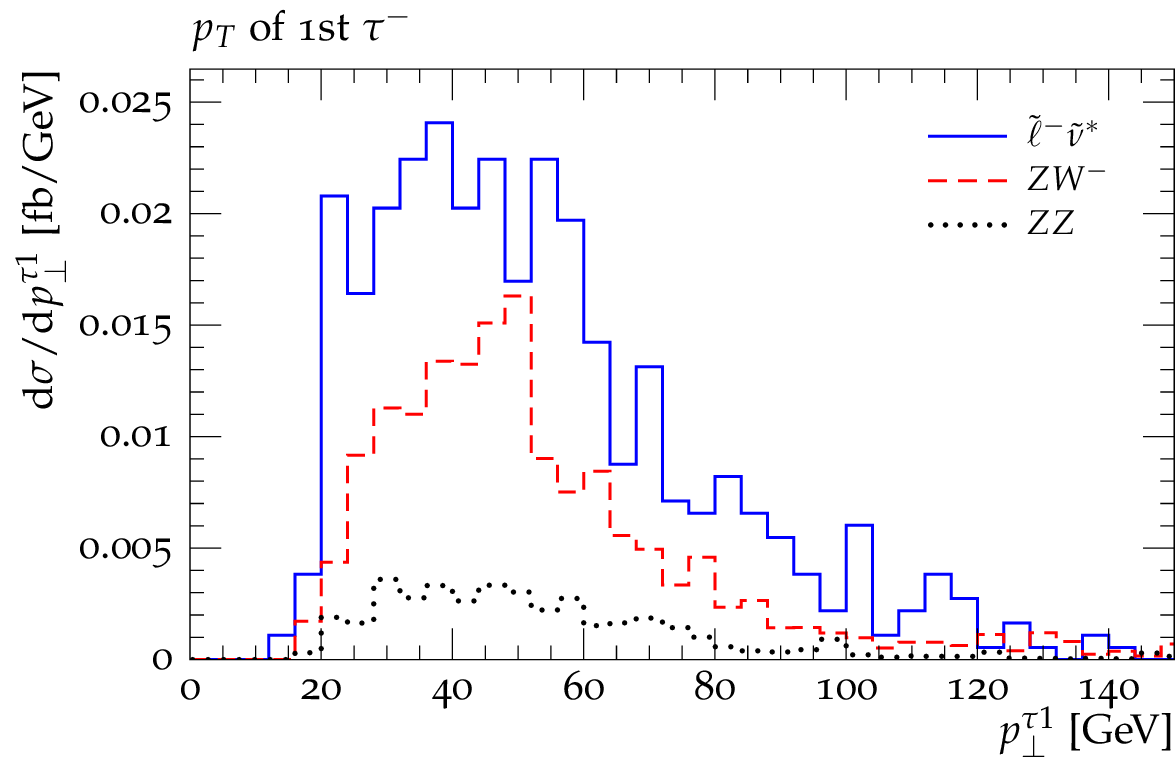}
\includegraphics[width=0.48\linewidth]{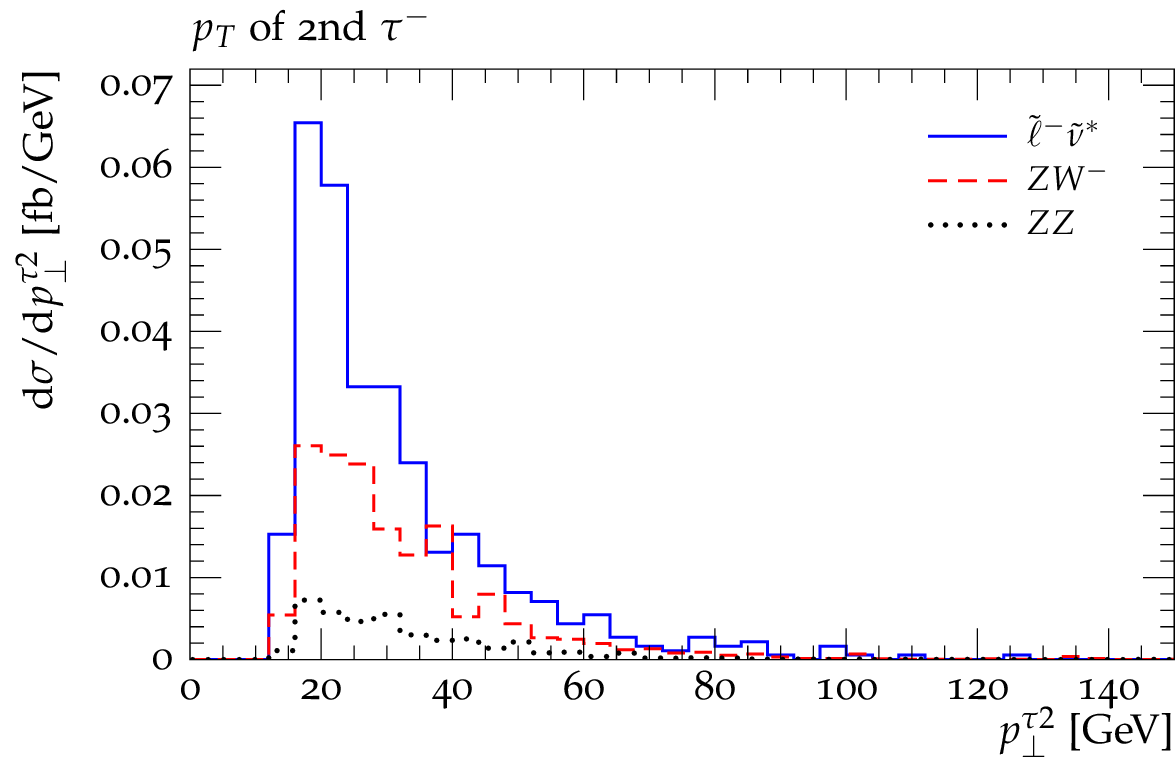}
\hfill (c) \hspace{7cm} (d) \hfill
\caption{The distribution in the (a,b) $\tau^{+}$ and (c,d) $\tau^{-}$ transverse momenta, $p_{T}^{\tau^{+}}$ and $p_{T}^{\tau^{-}}$ respectively. Channels giving negligible contribution are not shown here. }\label{Fig:PT-tau}
\end{center}
\end{figure}

The transverse momentum distributions of the two taus are shown in
Fig.~\ref{Fig:PT-tau}. Here $\tau_1$ is the hardest tau and $\tau_2$
is the softer tau. The hardest tau momentum peaks at around 40~GeV
and the $p_T^{\tau_1}$ cut (in the basic cuts) does not reduce the
signal much. For the softer tau, however, the cut is significant.
Increasing the $p_T^{\tau_2}$ cut from 10~GeV to 30~GeV, for
example, can reduce the effective cross section by $\sim 40$\%. We
choose our cut at 15~GeV, which although difficult from experimental
point of view should be possible at the LHC. Again, we see that it is
difficult to reduce the $ZW$ background any further.

\begin{figure}[htbp]
\begin{center}
\includegraphics[width=0.48\linewidth]{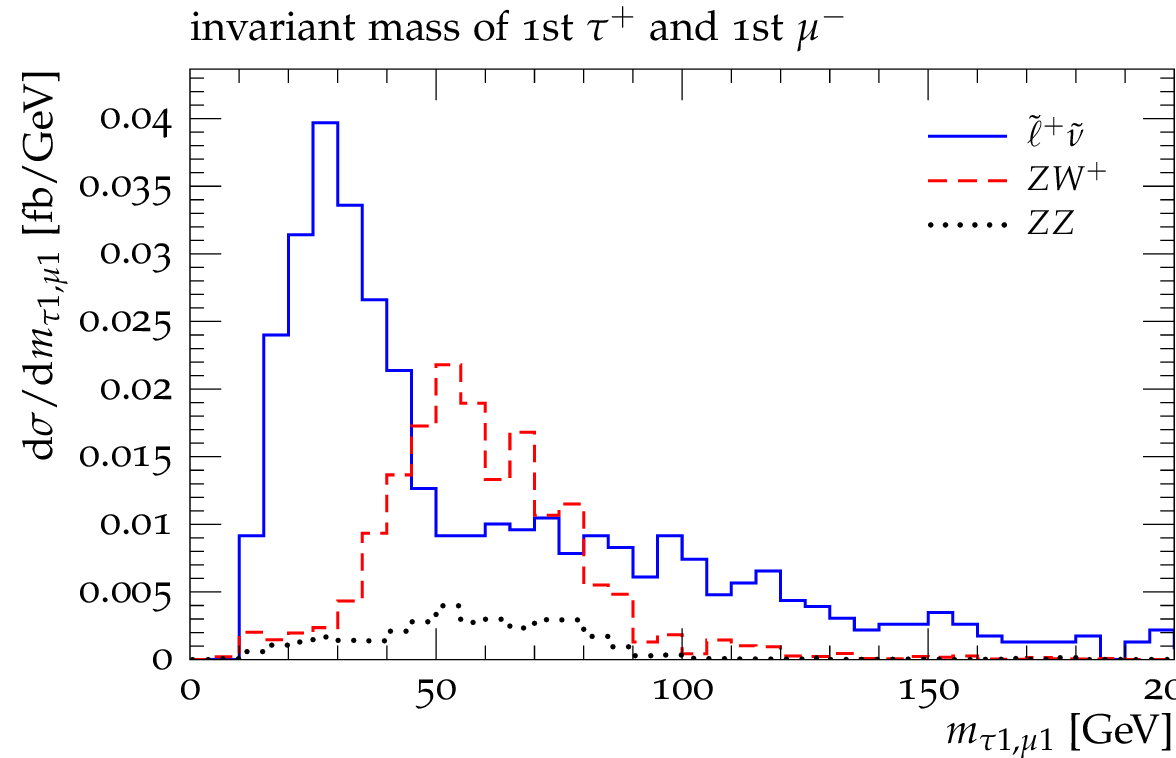}
\includegraphics[width=0.48\linewidth]{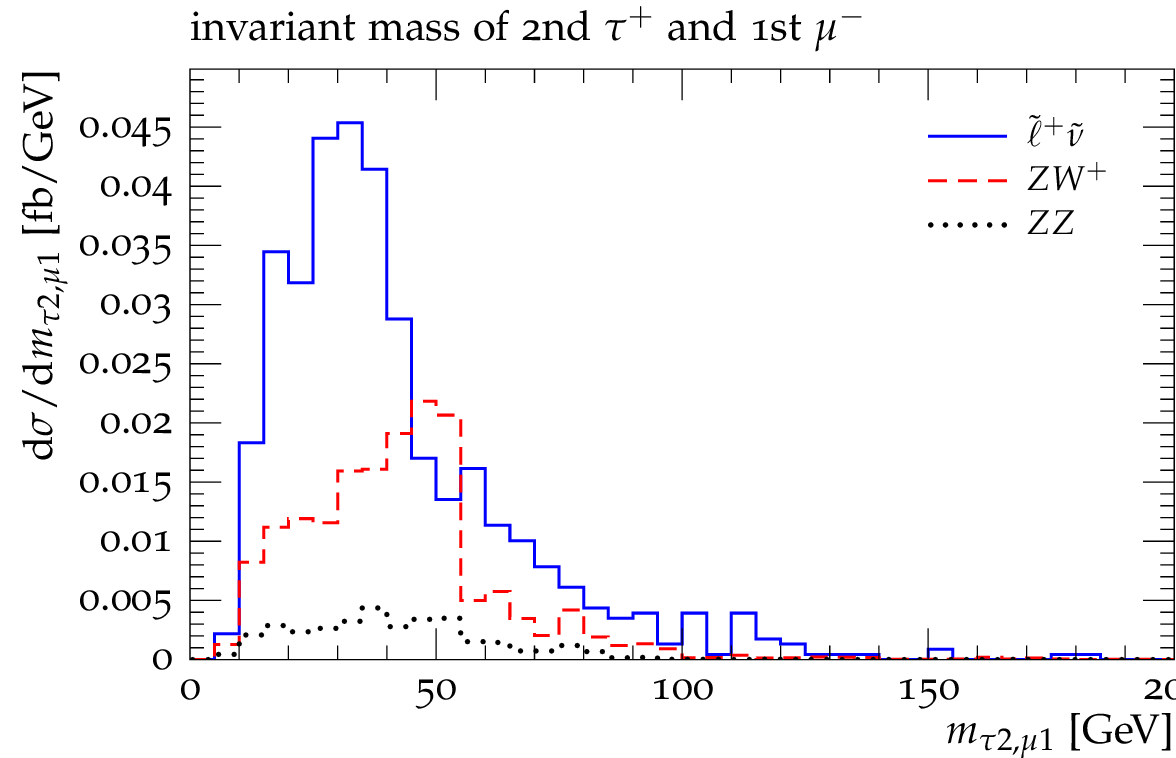}
\hfill (a) \hspace{7cm} (b) \hfill
\vskip 0.1in
\includegraphics[width=0.48\linewidth]{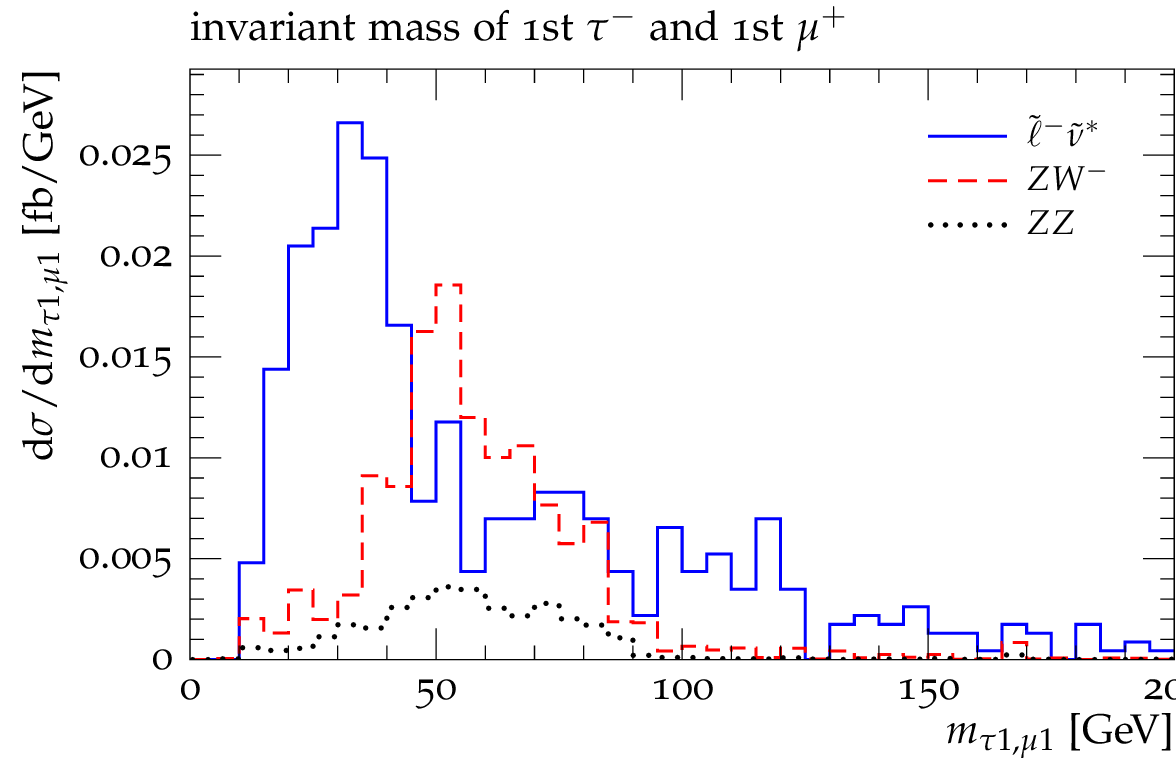}
\includegraphics[width=0.48\linewidth]{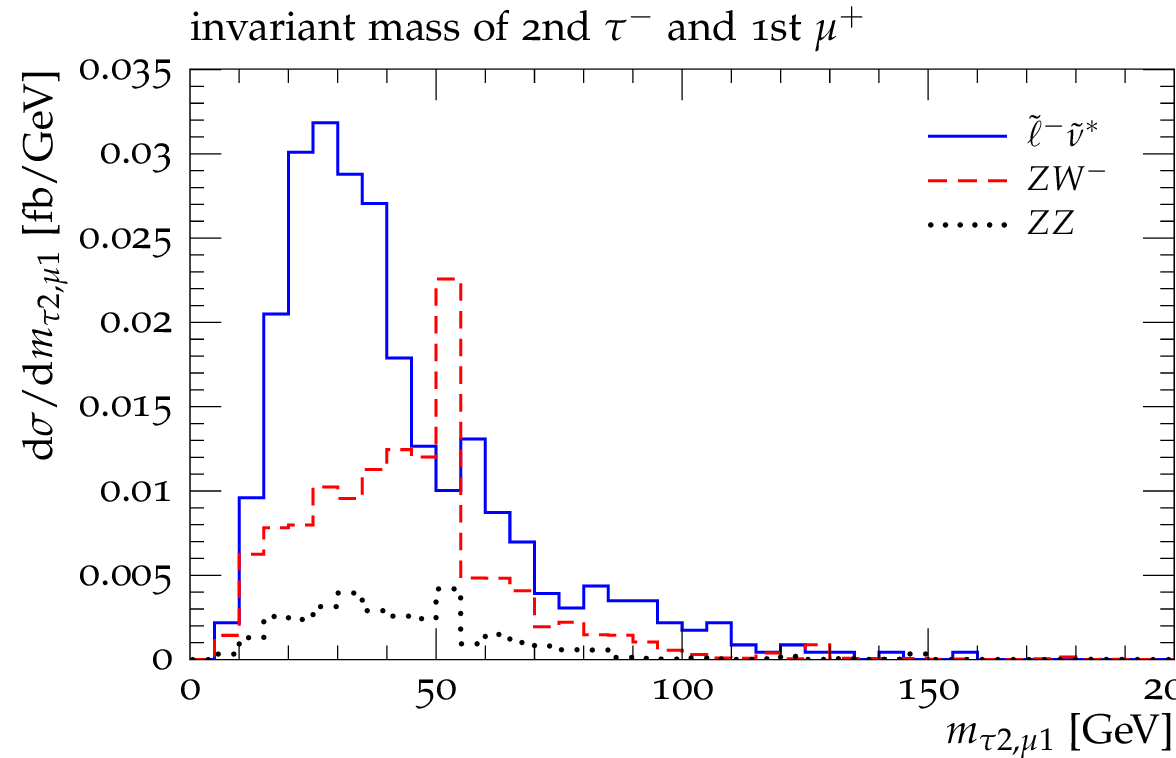}
\hfill (c) \hspace{7cm} (d) \hfill
\caption{The distribution in the (a,b) $\mu^{-} \tau^{+}$ and (c,d)
$\mu^{+} \tau^{-}$ invariant masses. Here $\tau_1$ is the hardest
tau, and $\tau_2$ is the second hardest. Channels giving negligible contribution are not shown here.}\label{Fig:M-mutau}
\end{center}
\end{figure}

%
Fig.~\ref{Fig:M-mutau} shows the distribution in the invariant mass
of $\mu$ and $\tau$ pair.  We note that the signal distribution is
concentrated at $m_{\tau\mu} < 50$~GeV. This suggest a bump feature
which arises from $\mu$ and $\tau$ pair coming from the same decay
chain (i.e.\ from $\tilde{\nu}_\mu$, see Eq.~\req{eq:snu-decay}).
The endpoint of this bump indicates a mass gap of $\sim 50$~GeV
between $\tilde{\nu}_\mu$ and $\tilde{\nu}_\tau$ which agrees with
our mass spectrum. There is, also, a smooth distribution without an
endpoint for high invariant masses. This arises from pairing the
$\mu$ with the $\tau$ that comes from smuon decay.  We might be able
to cut $ZW$ a little bit more in this case by cutting out around
60~GeV, but the gain is not significant.

\begin{figure}[htbp]
\begin{center}
\includegraphics[width=0.48\linewidth]{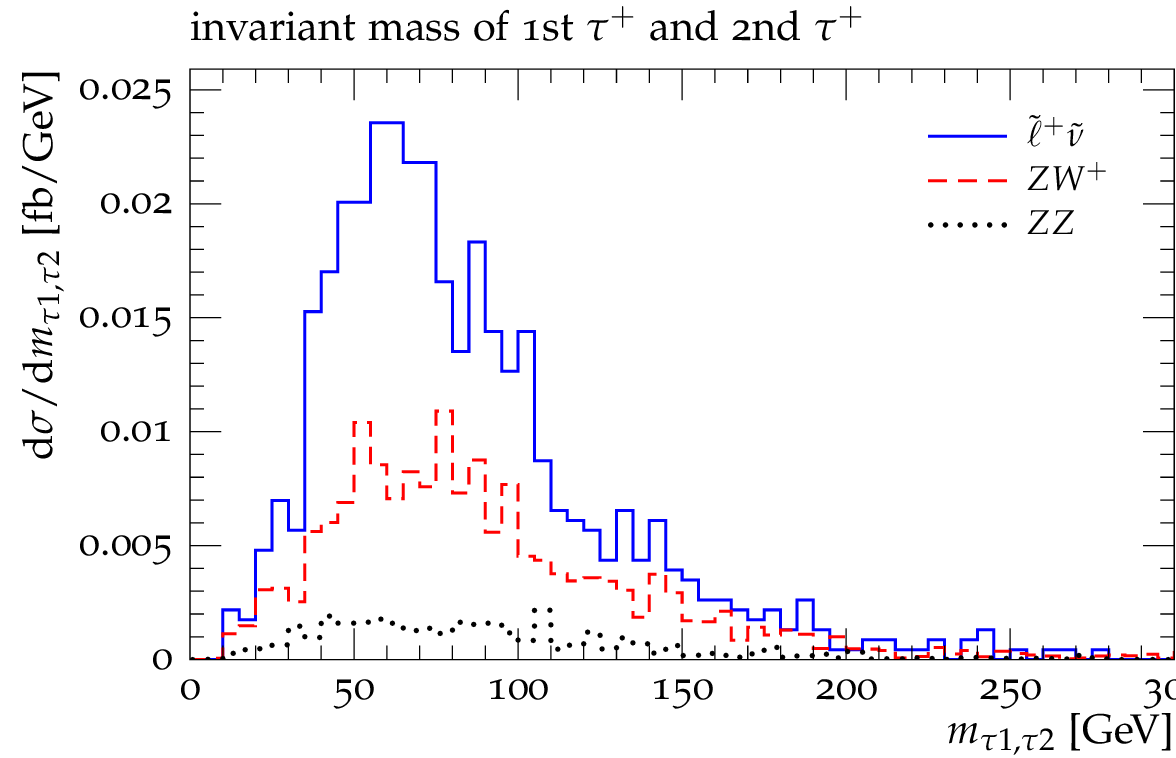}
\includegraphics[width=0.48\linewidth]{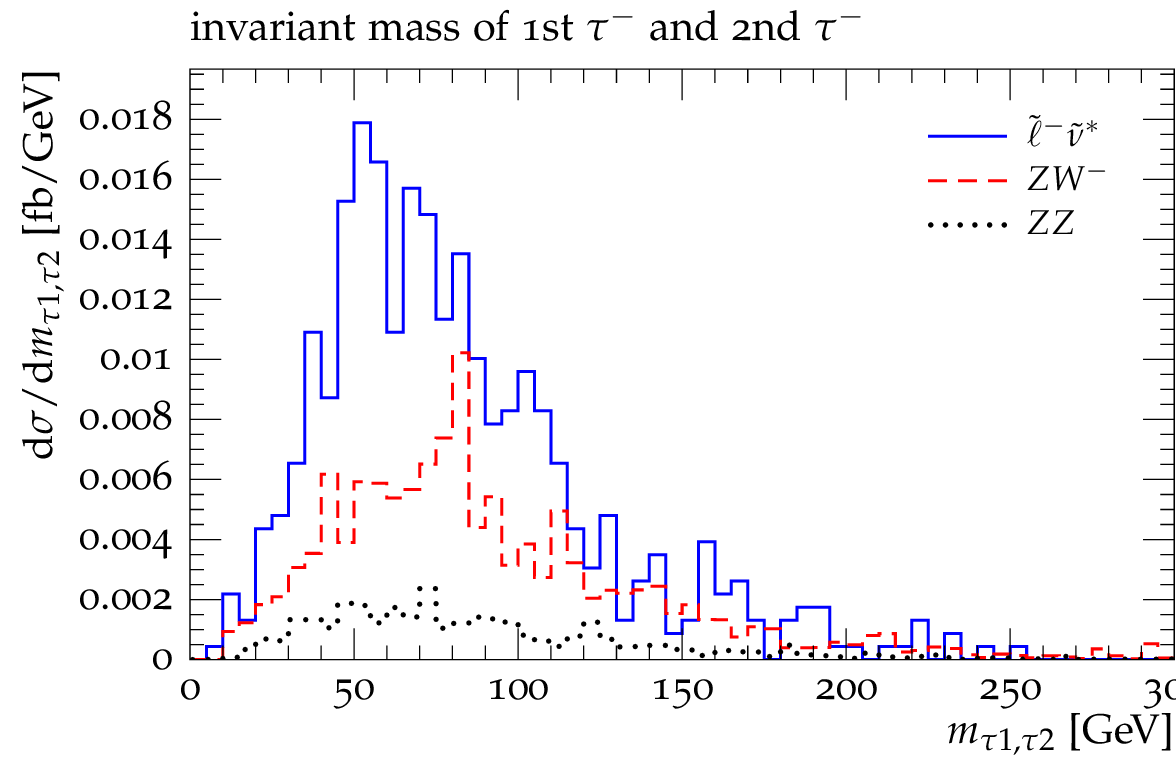}
\hfill (a) \hspace{7cm} (b) \hfill
\caption{The distribution in the (a) $\tau^{+} \tau^{+}$ and (b) $\tau^{-} \tau^{-}$ invariant masses. Channels giving negligible contribution are not shown here.} \label{Fig:M-tautau}
\end{center}
\end{figure}

In Fig.~\ref{Fig:M-tautau} we show the invariant mass distribution
of $\tau \tau$ pair. There is no endpoint feature seen in this plot,
suggesting that the two taus always come from the opposite decay
chains. On the other hand, notice that the invariant mass
distribution peaks at around 50~GeV, indicating that both taus are
coming from decays of weak scale particles.

In all of the plots above, the tau efficiency factor $\epsilon$ has
not been included. Note that since both signal and background are
affected by $\epsilon$, including this factor would only rescale the
distribution height but would not change the ratio between signal
and background~\footnote{Note, however, that $\epsilon$ varies with
respect to some observables such as the tau transverse momentum
$p_T^\tau$ and rapidity. Therefore the rescaling is not constant.},
except for the fake tau rate which is not included in the plots.
Here we assume that the fake tau rate, which depends on the real
data analysis, can be kept under control.

Significance can be estimated as follows
\begin{equation}
\frac{S}{\sqrt{S + B}} = \frac{\sigma_S}{\sqrt{\sigma_S + \sigma_B}} \cdot \epsilon_{\rm eff} \cdot \sqrt{\int {\cal L}}
\end{equation}
where $\epsilon_{\rm eff}$ is the effective tau identification
efficiency factor over the whole spectrum and $\int {\cal L}$ is an
integrated luminosity. Here we have used the fact that there are two
taus in our signal, and assumed (for simplicity) the same effective
efficiency factor $\epsilon_{\rm eff}$ for both taus. The tau charge
identification efficiency is implicitly included in $\epsilon_{\rm
eff}$, i.e. $\epsilon \equiv \epsilon_\tau \epsilon_{\rm charge}$.

Recalling the effective cross sections for the $\tau_h \tau_h \ell$
signal after the cuts as summarized in Table~\ref{Tab:trilep1}, we
find that the total effective cross sections are $\sigma_{\rm
opt.A}({\rm SUSY}) \simeq 3.2$~fb and $\sigma_{\rm opt.A}({\rm SM})
\simeq 1.9$~fb respectively, including both $\tau^+ \tau^+ \mu^-$
and $\tau^- \tau^- \mu^+$. We see that for 5-$\sigma$ discovery
level, the required integrated luminosity is
\begin{equation}
\int {\cal L} (5\sigma)_{\rm opt.A} \simeq 12.5/\epsilon_{\rm eff}^2 \ ({\rm fb}^{-1}) \, .
\end{equation}
Taking $\epsilon_{\rm eff} = 0.4$~\cite{Atlas-LHC}, for example, we
find that 5-$\sigma$ discovery requires about 80~fb$^{-1}$ of data.

On the other hand, for $p_T^{j}$ analysis (opt.B), we have
$\sigma_{\rm opt.B}({\rm SUSY}) \simeq 7.6$~fb and $\sigma_{\rm
opt.B}({\rm SM}) \simeq 0.7$~fb respectively, leading to
\begin{equation}
\int {\cal L} (5\sigma)_{\rm opt.B} \simeq 3.6/\epsilon_{\rm eff}^2 \ ({\rm fb}^{-1}) \, .
\end{equation}
Thus, with $\epsilon_{\rm eff} = 0.4$, 5-$\sigma$ level in this case
requires only about 23~fb$^{-1}$ of data. Note that this is not
necessarily the most promising channel to look for high-$p_T$ jets,
i.e. we should also  compare it with pure jets channel etc, each of
which requires a separate set of analysis.

Here, we have assumed 14~TeV CM energy for our analysis. Even using
the highest energy expected at the LHC we see that we need a
significant amount of data for discovery. We can deduce from
Table~\ref{prod-table} that it would be very difficult with 10~TeV
CM energy and practically impossible with 7~TeV.  Thus, we hope that
the LHC can overcome its technical difficulties and reach the
original designed energy of 14~TeV.

\section{Conclusion}

We have studied the leptonic signatures of a model in which
tau-sneutrino is the effectively stable lightest supersymmetric
particle at the LHC. The model that we consider has relatively heavy
charginos and neutralinos, and relatively light sleptons and
sneutrinos. The cross sections for pure leptonic signals are
generally small, partly due to the fact that neutralino-chargino
associated production in this model is suppressed by the heavy
gaugino masses. Nevertheless, we find that the trilepton signature
is still interesting to look at. It consists of a signal with two
like sign taus and one electron or muon of opposite sign, coming
from $\tilde{\ell}_L \tilde{\nu}_\ell$ production.

We employ an inclusive search strategy in generating signals and use
a set of cuts to look at this particular signature. At the Tevatron,
the sparticle production rates are too small to yield any observable
supersymmetric signal in our scenario. At the LHC, sufficiently
large CM energy is still required. With 14~TeV, and the optimized
cuts, we can obtain 5-$\sigma$ SUSY trilepton signals after $\sim
80$~fb$^{-1}$ integrated luminosity. We, also, investigated the
leptons + jets signatures, and noticed that we can use $p_T^j$ cut
to observe new physics signal above the standard model background.
In our case, for $p_T^{j_1} \gappeq 200$~GeV, where $j_1$ is the
hardest jet, SUSY QCD is dominant due to the large mass gap between
the squark sector and the slepton sector. In this way we can obtain
a significant signal-to-background ratio after 23~fb$^{-1}$ of
integrated luminosity. However, this does not tell us much about the
underlying model, since this is generally true for any
supersymmetric model with heavy squarks and gluino. Note that these
are optimistic estimates, assuming that we can suppress the fake-tau
event rate.

Our study suggests that the search for supersymmetry can be quite
challenging, depending on the specific supersymmetric model. This is
especially true when we want to look into the detailed
characteristics of the model. Even though the slepton spectrum is
relatively light, around 100~GeV, we still need large amount of data
and high energy to see a significant excess of signal over
background.

If this scenario is realized in nature, a big challenge in the data
analysis at the LHC would come from tau reconstruction and
identification, as well as rejection of fake taus. As tau can appear
copiously in many models beyond SM, we might need new methods in
this aspect. Indeed, there are ongoing efforts to alleviate these
problems~\cite{Nattermann:2009gh}. Nevertheless, hadron colliders,
such as the LHC, might not be sufficient to explore the physics
beyond the Standard Model. In this case, a lepton collider such as
the proposed $e^+e^-$ International Linear Collider (ILC)~\cite{ilc}
would help. For the ILC, the tau signal would be much cleaner. Only
then it would be possible to probe the model further, for example by
reconstructing masses and measuring couplings.

\section*{Acknowledgments}
\noindent We received a lot of feedback during this project. Our
special thank goes to Frank Krauss for many invaluable suggestions.
We also thank Steffen Schumann, Tanju Gleisberg, David Grellscheid,
Teruki Kamon, Bhaskar Dutta, Ayres Freitas, Gudrid Moortgat-Pick,
Peter Richardson, Andy Buckley, Alice Bean and Graham Wilson for
various information and useful discussions.
The work of YS was supported in part by the DOE  grant DE-FG02-04ER41308.
KR is supported by the EU Network MRTN-CT-2006-035505 ``Tools and Precision Calculations for Physics Discoveries at Colliders" (HEP-Tools).


\vskip 0.5in

\begin{thebibliography}{99}
\bibitem{LHC-ILC}
  G.~Weiglein {\it et al.}  [LHC/LC Study Group],
  Phys.\ Rept.\  {\bf 426} (2006) 47
  [arXiv:hep-ph/0410364].


\bibitem{EHNOS}
J. Ellis, J.S. Hagelin, D.V. Nanopoulos, K.A. Olive
and M. Srednicki, Nucl. Phys. B {\bf 238} (1984) 453; see also
H. Goldberg, Phys. Rev. Lett. {\bf 50} (1983) 1419.

\bibitem{Ellis:2003dn}
  J.~R.~Ellis, K.~A.~Olive, Y.~Santoso and V.~C.~Spanos,
  Phys.\ Lett.\  B {\bf 588} (2004) 7
  [arXiv:hep-ph/0312262].

\bibitem{FengGDM}
  J.~L.~Feng, S.~Su and F.~Takayama,
  Phys.\ Rev.\ D {\bf 70} (2004) 075019
  [arXiv:hep-ph/0404231];
  Phys.\ Rev.\ D {\bf 70} (2004) 063514
  [arXiv:hep-ph/0404198].


\bibitem{Covi:2007xj}
  L.~Covi and S.~Kraml,
  JHEP {\bf 0708} (2007) 015
  [arXiv:hep-ph/0703130].


\bibitem{Katz:2009qx}
  A.~Katz and B.~Tweedie,
  Phys.\ Rev.\  D {\bf 81} (2010) 035012
  [arXiv:0911.4132 [hep-ph]];
  Phys.\ Rev.\  D {\bf 81} (2010) 115003
  [arXiv:1003.5664 [hep-ph]].

\bibitem{Baer:1995nq}
  H.~Baer, C.~h.~Chen, F.~Paige and X.~Tata,
  Phys.\ Rev.\  D {\bf 52} (1995) 2746
  [arXiv:hep-ph/9503271].

\bibitem{Baer:1995va}
  H.~Baer, C.~h.~Chen, F.~Paige and X.~Tata,
  Phys.\ Rev.\  D {\bf 53} (1996) 6241
  [arXiv:hep-ph/9512383].

\bibitem{ourNUHM}
J.~R.~Ellis, K.~A.~Olive and Y.~Santoso,
  Phys.\ Lett.\ B {\bf 539} (2002) 107
  [arXiv:hep-ph/0204192];
J.~R.~Ellis, T.~Falk, K.~A.~Olive and Y.~Santoso,
  Nucl.\ Phys.\ B {\bf 652} (2003) 259
  [arXiv:hep-ph/0210205].

\bibitem{Ellis:2008as}
  J.~R.~Ellis, K.~A.~Olive and Y.~Santoso,
  JHEP {\bf 0810} (2008) 005
  [arXiv:0807.3736 [hep-ph]].

\bibitem{mtop1724}
    [Tevatron Electroweak Working Group, CDF and D0 Collaborations],
  arXiv:0808.1089 [hep-ex].

\bibitem{fh}
 S.~Heinemeyer, W.~Hollik and G.~Weiglein,
 Comput.\ Phys.\ Commun.\  {\bf 124} (2000) 76
 [arXiv:hep-ph/9812320];
 S.~Heinemeyer, W.~Hollik and G.~Weiglein,
 Eur.\ Phys.\ J.\ C {\bf 9} (1999) 343
 [arXiv:hep-ph/9812472].


\bibitem{Ambrosanio:1995az}
  S.~Ambrosanio and B.~Mele,
  Phys.\ Rev.\  D {\bf 53} (1996) 2541
  [arXiv:hep-ph/9508237].


\bibitem{sdecay}
  M.~Muhlleitner, A.~Djouadi and Y.~Mambrini,
  Comput.\ Phys.\ Commun.\  {\bf 168} (2005) 46
  [arXiv:hep-ph/0311167].


\bibitem{feynarts}
  J.~K\"{u}blbeck, M.~Bohm and A.~Denner,
  Comput.\ Phys.\ Commun.\ {\bf 60} (1990) 165; T.~Hahn,
  Comput.\ Phys.\ Commun.\ {\bf 140} (2001) 418
  [arXiv:hep-ph/0012260v2];
  T.~Hahn and M.~Perez-Victoria,
  Comput.\ Phys.\ Commun.\ {\bf 118} (1999) 153
  [arXiv:hep-ph/9807565];
  J.~A.~M.~Vermaseren,
  arXiv:math-ph/0010025;
  T.~Hahn and C.~Schappacher,
  Comput.\ Phys.\ Commun.\ {\bf 143} (2002) 54 [arXiv:hep-ph/0105349].

\bibitem{Kraml:2007sx}
  S.~Kraml and D.~T.~Nhung,
  JHEP {\bf 0802} (2008) 061
  [arXiv:0712.1986 [hep-ph]].

\bibitem{Herwig}
  M.~Bahr {\it et al.},
  Eur.\ Phys.\ J.\  C {\bf 58} (2008) 639
  [arXiv:0803.0883 [hep-ph]].





\bibitem{SPS}
  B.~C.~Allanach {\it et al.},
in {\it Proc. of the APS/DPF/DPB Summer Study on the Future of Particle Physics (Snowmass 2001) } ed. N.~Graf,
  Eur.\ Phys.\ J.\  C {\bf 25} (2002) 113
  [arXiv:hep-ph/0202233].

\bibitem{Alwall:2007ed}
  J.~Alwall, D.~Rainwater and T.~Plehn,
  Phys.\ Rev.\  D {\bf 76} (2007) 055006
  [arXiv:0706.0536 [hep-ph]].

\bibitem{Martin:1999ww}
  A.~D.~Martin, R.~G.~Roberts, W.~J.~Stirling and R.~S.~Thorne,
  Eur.\ Phys.\ J.\  C {\bf 14} (2000) 133
  [arXiv:hep-ph/9907231].

\bibitem{Frixione:1992pj}
  S.~Frixione, P.~Nason and G.~Ridolfi,
  Nucl.\ Phys.\  B {\bf 383} (1992) 3.

\bibitem{Campbell:1999ah}
  J.~M.~Campbell and R.~K.~Ellis,
  Phys.\ Rev.\  D {\bf 60} (1999) 113006
  [arXiv:hep-ph/9905386].

\bibitem{Mele:1990bq}
  B.~Mele, P.~Nason and G.~Ridolfi,
  Nucl.\ Phys.\  B {\bf 357} (1991) 409.
  L.~J.~Dixon, Z.~Kunszt and A.~Signer,
  Phys.\ Rev.\  D {\bf 60} (1999) 114037
  [arXiv:hep-ph/9907305].

\bibitem{Frixione:1993yp}
  S.~Frixione,
  Nucl.\ Phys.\  B {\bf 410} (1993) 280;
  T.~Binoth, M.~Ciccolini, N.~Kauer and M.~Kramer,
  JHEP {\bf 0612} (2006) 046
  [arXiv:hep-ph/0611170].





\bibitem{CDF-trilep}
  T.~Aaltonen {\it et al.}  [CDF Collaboration],
  Phys.\ Rev.\ Lett.\  {\bf 101} (2008) 251801
  [arXiv:0808.2446 [hep-ex]].


\bibitem{3lep}
  H.~Baer, C.~h.~Chen, F.~Paige and X.~Tata,
  Phys.\ Rev.\  D {\bf 50} (1994) 4508
  [arXiv:hep-ph/9404212];
  V.~D.~Barger, C.~Kao and T.~j.~Li,
  Phys.\ Lett.\  B {\bf 433} (1998) 328
  [arXiv:hep-ph/9804451];
  V.~D.~Barger and C.~Kao,
  Phys.\ Rev.\  D {\bf 60} (1999) 115015
  [arXiv:hep-ph/9811489];
  H.~Baer, M.~Drees, F.~Paige, P.~Quintana and X.~Tata,
  Phys.\ Rev.\  D {\bf 61} (2000) 095007
  [arXiv:hep-ph/9906233];
  E.~Accomando, R.~L.~Arnowitt and B.~Dutta,
  Phys.\ Lett.\  B {\bf 475} (2000) 176
  [arXiv:hep-ph/9811300];
  Z.~Sullivan and E.~L.~Berger,
  Phys.\ Rev.\  D {\bf 78} (2008) 034030
  [arXiv:0805.3720 [hep-ph]].


\bibitem{4lep-via-H}
  G.~Bian, M.~Bisset, N.~Kersting, Y.~Liu and X.~Wang,
  Eur.\ Phys.\ J.\  C {\bf 53} (2008) 429
  [arXiv:hep-ph/0611316];
  P.~Huang, N.~Kersting and H.~H.~Yang,
  Phys.\ Rev.\  D {\bf 77} (2008) 075011
  [arXiv:0801.0041 [hep-ph]].


\bibitem{tripleWZ}
  F.~Campanario, V.~Hankele, C.~Oleari, S.~Prestel and D.~Zeppenfeld,
  Phys.\ Rev.\  D {\bf 78} (2008) 094012
  [arXiv:0809.0790 [hep-ph]].



\bibitem{Atlas-LHC}
  G.~Aad {\it et al.}  [ATLAS Collaboration],
  JINST {\bf 3} (2008) S08003.

\bibitem{CMS-TDR}
  G.~L.~Bayatian {\it et al.}  [CMS Collaboration],
  J.\ Phys.\ G {\bf 34} (2007) 995.





\bibitem{CMS-tau}
  G.~Bagliesi {\it et al.}, CERN-CMS-NOTE-2006-028.

\bibitem{ATLAS-tau}
  A.~F.~Saavedra  [ATLAS Collaboration], ATL-PHYS-PROC-2009-007.


\bibitem{W2j}
  J.~M.~Campbell, R.~K.~Ellis and D.~L.~Rainwater,
  Phys.\ Rev.\  D {\bf 68} (2003) 094021
  [arXiv:hep-ph/0308195].

\bibitem{sherpa}
  T.~Gleisberg, S.~Hoche, F.~Krauss, M.~Schonherr, S.~Schumann, F.~Siegert and J.~Winter,
  JHEP {\bf 0902} (2009) 007
  [arXiv:0811.4622 [hep-ph]].

\bibitem{comix}
  T.~Gleisberg and S.~Hoche,
  JHEP {\bf 0812} (2008) 039
  [arXiv:0808.3674 [hep-ph]].


\bibitem{antikTalg}
  M.~Cacciari, G.~P.~Salam and G.~Soyez,
  JHEP {\bf 0804} (2008) 063
  [arXiv:0802.1189 [hep-ph]].


\bibitem{rivet}
  A.~Buckley {\it et al.},
  arXiv:1003.0694 [hep-ph].

\bibitem{fastjet}
  M.~Cacciari and G.~P.~Salam,
  Phys.\ Lett.\  B {\bf 641} (2006) 57
  [arXiv:hep-ph/0512210]; M. Cacciari, G.P. Salam and G. Soyez, \texttt{http://fastjet.fr}

\bibitem{Baer:2007ya}
  H.~Baer, V.~Barger, G.~Shaughnessy, H.~Summy and L.~t.~Wang,
  Phys.\ Rev.\  D {\bf 75} (2007) 095010
  [arXiv:hep-ph/0703289].

\bibitem{Baer:2008kc}
  H.~Baer, H.~Prosper and H.~Summy,
  Phys.\ Rev.\  D {\bf 77} (2008) 055017
  [arXiv:0801.3799 [hep-ph]].


\bibitem{Nattermann:2009gh}
  T.~Nattermann, K.~Desch, P.~Wienemann and C.~Zendler,
  JHEP {\bf 0904} (2009) 057
  [arXiv:0903.0714 [hep-ph]].

\bibitem{ilc}
  J.~Brau {\it et al.},
  ILC-REPORT-2007-001, CERN-2007-006 (2007);
  A.~Djouadi, J.~Lykken, K.~Monig, Y.~Okada, M.~J.~Oreglia and S.~Yamashita,
  arXiv:0709.1893 [hep-ph].

\end{thebibliography}
\end{document}